\documentclass[british]{article}
\usepackage{charter}

\usepackage[T1]{fontenc}
\usepackage[latin9]{inputenc}
\usepackage{geometry}
\usepackage{babel}
\usepackage{textcomp}
\usepackage{amsmath}
\usepackage{amssymb}
\usepackage{graphicx}
\usepackage{esint}
\usepackage[authoryear]{natbib}
\usepackage[unicode=true,pdfusetitle,
 bookmarks=true,bookmarksnumbered=false,bookmarksopen=false,
 breaklinks=false,pdfborder={0 0 1},backref=section,colorlinks=false]
 {hyperref}
\hypersetup{
 hidelinks}

\makeatletter
\usepackage{babel}

\@ifundefined{showcaptionsetup}{}{%
 \PassOptionsToPackage{caption=false}{subfig}}
\usepackage{subfig}
\makeatother

\begin{document}

\author{Simon Barthelmé \\
Psychology, University of Geneva. Boulevard du Pont-d'Arve 40 1205
Genève, Switzerland. \and Hans Trukenbrod, Ralf Engbert \\
 Psychology, University of Potsdam Karl-Liebknecht-Str. 24-25 14476
Potsdam OT Golm, Germany \and Felix Wichmann \\
 1. Neural Information Processing Group, Faculty of Science, University
of Tübingen, \\
 Sand 6, 72076 Tübingen, Germany \\
 2. Bernstein Center for Computational Neuroscience Tübingen, \\
 Otfried-Müller-Str. 25, 72076 Tübingen, Germany \\
 3. Max Planck Institute for Intelligent Systems, Empirical Inference
Department, \\
 Spemannstr. 38, 72076 Tübingen, Germany}

\title{Modelling fixation locations using spatial point processes}
\maketitle
\begin{abstract}
Whenever eye movements are measured, a central part of the analysis
has to do with \emph{where }subjects fixate, and \emph{why} they fixated
where they fixated. To a first approximation, a set of fixations can
be viewed as a set of points in space: this implies that fixations
are spatial data and that the analysis of fixation locations can be
beneficially thought of as a spatial statistics problem. We argue
that thinking of fixation locations as arising from \emph{point processes
}is a very fruitful framework for eye movement data, helping turn
qualitative questions into quantitative ones.

We provide a tutorial introduction to some of the main ideas of the
field of spatial statistics, focusing especially on spatial Poisson
processes. We show how point processes help relate image properties
to fixation locations. 
In particular we show how point processes naturally express the idea
that image features' predictability for fixations may vary from one
image to another. We review other methods of analysis used in the
literature, show how they relate to point process theory, and argue
that thinking in terms of point processes substantially extends the
range of analyses that can be performed and clarify their interpretation. 
\end{abstract}
\global\long\def\bt{\bm{\theta}}
 \global\long\def\bmu{\bm{\mu}}
 \global\long\def\bl{\bm{\lambda}}
 \global\long\def\be{\bm{\eta}}
 \global\long\def\E{\mathbb{E}}
 \global\long\def\by{\bm{y}}
 \global\long\def\ys{\bm{y}^{\star}}
 \global\long\def\bS{\bm{\Sigma}}
 \global\long\def\N{\mathcal{N}}
 \global\long\def\I{\mathbb{I}}
 \global\long\def\R{\mathbb{R}}
 \global\long\def\bb{{\bf \beta}}

Eye movement recordings are some of the most complex data available
to behavioural scientists. At the most basic level they are long sequences
of measured eye positions, a very high dimensional signal containing
saccades, fixations, micro-saccades, drift, and their myriad variations
\citep{CiuffredaTannen:EyeMovementBasics}. There are already many
methods that process the raw data and turn it into a more manageable
format, checking for calibration, distinguishing saccades from other
eye movements (e.g., \citealp{EngbertMergenthaler:MicrosaccadesTriggRetImSlip,MergenthalerEngbert:MicrosaccDiffSaccScenePerception}).
Our focus is rather on fixation locations.

In the kind of experiment that will serve as an example throughout
the paper, subjects were shown a number of pictures on a computer
screen, under no particular instructions. The resulting data are a
number of points in space, representing what people looked at in the
picture---the fixation locations. The fact that fixations tend to
cluster shows that people favour certain locations and do not simply
explore at random. Thus one natural question to ask is why are certain
locations preferred? 

We argue that a very fruitful approach to the problem is to be found
in the methods of spatial statistics \citep{Diggle:StatisticalAnalysisSpatialPointPatterns,Illian:StatAnalysisSpatPointPatterns}.
A sizeable part of spatial statistics is concerned with how things
are distributed in space, and fixations are ``things'' distributed
in space. We will introduce the concepts of point processes and latent
fields, and explain how these can be applied to fixations. We will
show how this lets us put the important (and much researched) issue
of low-level saliency on firmer statistical ground. We will begin
with simple models and gradually build up to more sophisticated models
that attempt to separate the various factors that influence the location
of fixations and deal with non-stationarities.

Using the point process framework, we replicate results obtained previously
with other methods, but also show how the basic tools of point process
models can be used as building blocks for a variety of data analyses.
They also help shed new light on old tools, and we will argue that
classical methods based on analysing the contents of image patches
around fixated locations make the most sense when seen in the context
of point process models.

\section{Analysing eye movement data}

\subsection{Eye movements}

While looking at a static scene our eyes perform a sequence of rapid
jerk-like movements (saccades) interrupted by moments of relative
stability (fixations)%
\footnote{Our eyes are never perfectly still and miniature eye movements (microsaccades,
drift, tremor) can be observed during fixations \citep{CiuffredaTannen:EyeMovementBasics}.%
}. One reason for this fixation-saccade strategy arises from the inhomogeneity
of the visual field \citep{Land.BOOK.2009}. Visual acuity is highest
at the center of gaze, i.e. the fovea (within 1\textdegree{}~eccentricity),
and declines towards the periphery as a function of eccentricity.
Thus saccades are needed to move the fovea to selected parts of an
image for high resolution analysis. About 3--4 saccades are generated
each second. An average saccade moves the eyes 4--5\textdegree{}~during
scene perception and, depending on the amplitude, lasts between 20--50~ms.
Due to saccadic suppression (and the the high velocity of saccades)
vision is hampered during saccades \citep{Matin.PsycholBull.1974}
and information uptake is restricted to the time in between, i.e.\ the
fixations. During a fixation gaze is on average held stationary for
250--300~ms but individual fixation durations are highly variable
and range from less than a hundred milliseconds to more than a second.
For recent reviews on eye movements and eye movements during scene
perception see \citet{Rayner.QJExpPsychol.2009} and \citet{Henderson.Liversedge.2011},
respectively.

During scene perception fixations cluster on ``informative'' parts
of an image whereas other parts only receive few or no fixations.
This behavior has been observed between and within observers and has
been associated with several factors. Due to the close coupling of
stimulus features and attention \citep{Wolfe.NatRevNeurosci.2004}
as well as eye movements and attention \citep{Deubel.VisionRes.1996},
local image features like contrast, edges, and color are assumed to
guide eye movements. In their influential model of visual saliency,
\citet{IttiKoch:ComputationalModellingVisualAttention} combine several
of these factors to predict fixation locations. However, rather simple
calculations like edge detectors \citep{TatlerVincent:BehavBiasesEyeGuidance}
or center-surround patterns combined with contrast-gain control \citep{Kienzle:CenterSurroundPatternsOptimalPredictors}
seem to predict eye movements similarly well. The saliency approach
has generated a lot of interest in research on 
the prediction of fixation locations and has led to the development
of a broad variety of different models. A recent summary can be found
in \citet{Borji.IEEETPatternAnal.2013}.


Besides of local image features, fixations seem to be guided by faces,
persons, and objects \citep{Cerf.AdvNeuralInfoProcSyst.2008,Judd.CompVis.2009}.
Recently it has been argued that objects may be, on average, more
salient than scene background \citep{Einhauser.JVis.2008,Nuthmann:ObjectBasedAttentionalSelection}
suggesting that saccades might primarily target objects and that the
relation between objects, visual saliency and salient local image
features is just correlative in nature. The inspection behavior of
our eyes is further modulated by specific knowledge about a scene
acquired during the last fixations or more general knowledge acquired
on longer time scales \citep{Henderson.Henderson.2004}. Similarly,
the same image viewed under differing instructions changes the distribution
of fixation locations considerably \citep{Yarbus.BOOK.1967}. To account
for top-down modulations of fixation locations at a computational
level, \citet{Torralba:ContextualGuidanceEyeMovements} weighted a
saliency map with a-priori knowledge about a scene. Finally, the spatial
distribution of fixations is affected by factors independent of specific
images. \citet{Tatler:CentralFixationBias}, for example, reported
a strong bias towards central parts of an image. In conventional photographs
the effect may largely be caused by the tendency of photographers
to place interesting objects in the image center but, importantly,
the center bias remains in less structured images. 

Eye movements during scene perception have been a vibrant research
topic over the past years and the preceding paragraphs provide only
a brief overview of the diverse factors that contribute to the selection
of fixation locations. We illustrate the logic of spatial point processes
by using two of these factors in the upcoming sections: local image
properties---visual saliency---and the center bias. The concept of
point processes can easily be extended to more factors and helps to
assess the impact of various factors on eye movement control.

\subsection{Relating fixation locations to image properties}

\label{sub:FixLocationLocalProp} There is already a rather large
literature relating local image properties to fixation locations,
and it has given rise to many different methods for analysing fixation
locations. Some analyses are mostly descriptive, and compare image
content at fixated and non-fixated locations. Others take a stronger
modelling stance, and are built around the notion of a saliency map
combining a range of interesting image features. Given a saliency
map, one must somehow relate it to the data, and various methods have
been used to check whether a given saliency map has something useful
to say about eye movements. In this section we review the more popular
of the methods in use. As we explain below, in Section \ref{sub:Relating-patch-statistics-to-PP-theory}
, the point process framework outlined here helps to unify and make
sense of the great variety of methods in the field.

\citet{ReinagelZador:NatSceneStatsCenterGaze} had observers view
a set of natural images for a few seconds while their eye movements
were monitored. They selected image patches around gaze points, and
compared their content to that of control patches taken at random
from the images. Patches extracted around the center of gaze had higher
contrast and were less smooth than control patches. Reinagel and Zador's
work set a blueprint for many follow-up studies, such as \citet{Krieger:ObjectSceneAnalysisSaccadicEyeMovements}
and \citet{ParkhurstNiebur:SceneContentSelectedActiveVision}, although
they departed from the original by focusing on \emph{fixated }vs \emph{non-fixated
}patches. Fixated points are presumably the points of higher interest
to the observer, and to go from one to the next the eye may travel
through duller landscapes. Nonetheless the basic analysis pattern
remained: one compares the contents of selected patches to that of
patches drawn from random control locations.

Since the contents of the patches (e.g. their contrast) will differ
both within-categories and across, what one typically has is a distribution
of contrast values in fixated and control patches. The question is
whether these distributions differ, and asking whether the distributions
differ is mathematically equivalent to asking whether one can guess,
based on the contrast of a patch, whether the patch comes from the
fixated or the non-fixated set. We call this problem patch classification,
and we show in Section \ref{sub:The-patch-classification-problem}
that it has close ties to point process modelling---indeed, certain
forms of patch classification can be seen as approximations to point
process modelling.

The fact that fixated patches have distinctive local statistics could
suggest that it is exactly these distinctive local statistics that
attract gaze to a certain area. Certain models adopt this viewpoint
and assume that the visual system computes a bottom-up saliency map
based on local image features. The bottom-up saliency map is used
by the visual system (along with top-down influences) to direct the
eyes \citep{KochUllman:ShiftsInSelectiveVisualAttention}. Several
models of bottom-up saliency have been proposed (for a complete list
see \citealp{Borji.IEEETPatternAnal.2013}), based either on the architecture
of the visual cortex \citep{IttiKoch:ComputationalModellingVisualAttention}
or on computational considerations (e.g., \citealp{Kanan:SUNTopDownSaliencyUsingNaturalStats}),
but their essential feature for our purposes is that they take an
image as input and yield as output a saliency map. The computational
mechanism that produces the saliency map should ideally work out from
the local statistics of the image which areas are more visually conspicuous
and give them higher saliency scores. The model takes its validity
from the correlation between the saliency maps it produces and actual
eye movements.

How one goes from a saliency map to a set of eye movements is not
obvious, and \citet{Wilming:MeasuresLimitsModelsFixationSelection}
have found in their extensive review of the literature as many as
8 different performance measures. One solution is to look at area
counts \citep{Torralba:ContextualGuidanceEyeMovements}: if we pick
the 20\% most salient pixels in an image, they will define an area
that takes up 20\% of the picture. If much more than 20\% of the recorded
fixations are in this area, it is reasonable to say that the saliency
model gives us useful information, because by chance we'd expect this
proportion to be around 20\%.

A seemingly completely different solution is given by the very popular
AUC measure \citep{Tatler:VisualCorrelatesFixSel}, which uses the
patch classification viewpoint: fixated patches should have higher
salience than control patches. The situation is analoguous to a signal
detection paradigm: correctly classifying a patch as fixated is a
Hit, incorrectly classifying a patch as fixated is a False Alarm,
etc. A good saliency map should give both a high Hit Rate and a low
rate of False Alarms, and therefore performance can be quantified
by the area under the ROC curve (AUC): the higher the AUC, the better
the model.

One contribution of the point process framework is that we can prove
these two measures are actually tightly related, even though they
are rather different in origin (Section \ref{sub:ROC-and-Area-counts}).
There are many other ways to relate stimulus properties to fixation
locations, based for example on scanpaths \citep{Henderson:VisualSaliencyDoesNotAccountForEyeMovements},
on the number of fixations before entering a region of interest \citep{Underwood:EyeMovementsDuringSceneInspection},
on the distance between fixations and landmarks \citep{Mannan:RelationshipLocationSpatialFeaturesFixations},
etc. We cannot attempt here a complete unification of all measures,
but we hope to show that our proposed spatial point process framework
is general enough that such unification is at least theoretically
possible. In the next section we introduce the main ideas behind spatial
point-process models.

\section{Point processes}

We begin with a general overview on the art and science of generating
random sets of points in space. It is important to emphasise at this
stage that the models we will describe are entirely \emph{statistical
}in nature and not mechanistic: they do not assume anything about
how saccadic eye movements are generated by the brain \citep{Sparks:BrainstemControlSaccEyeMovements}.
In this sense they are more akin to linear regression models than,
e.g., biologically-inspired models of overt attention during reading
or visual search \citep{Engbert:SWIFTSaccadeGenDuringReading,Zelinsky:TheoryEyeMovementsTargetAcquisition}.
The goal of our modelling is to provide statistically sound and useful
summaries and visualizations of data, rather than come up with a full
story of how the brain goes about choosing where to allocate the next
saccade. What we lose in depth, we gain in generality, however: the
concepts that are highlighted here are applicable to the vast majority
of experiments in which fixations locations are of interest.

\subsection{Definition and examples}

In statistics, point patterns in space are usually described in terms
of point processes, which represent realisations from probability
distributions over sets of points. Just like linear regression models,
point processes have a deterministic and a stochastic component. In
linear models, the deterministic component describes the average value
of the dependent variable as a function of the independent ones, and
the stochastic component captures the fact that the model cannot predict
perfectly the value of the independent variable, for example because
of measurement noise. In the same way, point processes will have a
latent \emph{intensity function}, which describes the expected number
of points that will be found in a certain area, and a stochastic part
which captures prediction error and/or intrinsic variability.

We focus on a certain class of point process models known as inhomogeneous
Poisson processes. Some specific examples of inhomogeneous Poisson
processes should be familiar to most readers. These are temporal rather
than spatial, which means they generate random point sets in time
rather than in space, but equivalent concepts apply in both cases.

In neuroscience, Poisson processes are often used to characterize
neuronal spike trains (see e.g., \citealp{AbbottDayan}). The assumption
is that the number of spikes produced by a neuron in a given time
interval follows a Poisson distribution: for example, repeated presentation
of the same visual stimulus will produce a variable number of spikes,
but the variability will be well captured by a Poisson distribution.
Different stimuli will produce different average spike rates, but
spike rate will also vary over \emph{time }during the course of a
presentation, for example rising fast at stimulation onset and then
decaying. A useful description, summarized in Figure \ref{fig:Example-Temporal-IPP},
is in terms of a latent intensity function $\lambda(t)$ governing
the expected number of spikes observed in a certain time window. Formally,
$\int_{\tau}^{\tau+\delta}\lambda\left(t\right)\mbox{d}t$ gives the
expected number of spikes between times $\tau$ and $\tau+\delta$.
If we note $\mathbf{t}=t_{1},t_{2},\ldots,t_{k}$ the times at which
spikes occurred on a given trial, then $\mathbf{t}$ follows a inhomogeneous
Poisson Process (from now on IPP) distribution if, for all intervals
$\left(\tau,\tau+\delta\right)$, the number of spikes occurring in
the interval follows a Poisson distribution (with mean given by the
integral of $\lambda\left(t\right)$ over the interval).

The temporal IPP therefore gives us a distribution over sets of points
in time (in Figure \ref{fig:Example-Temporal-IPP}, over the interval
$[0,1]$). Extending to the spatial case is straightforward: we simply
define a new intensity function $\lambda(x,y)$ over space, and the
IPP now generates point sets such that the expected number of points
to appear in a certain area $A$ is $\int_{A}\lambda\left(x,y\right)\mbox{d}x\mbox{d}y$,
with the actual quantity again following a Poisson distribution. The
spatial IPP is illustrated on Figure \ref{fig:Example-Spatial-IPP}.

\begin{center}
\begin{figure}
\begin{centering}
\includegraphics[scale=0.7]{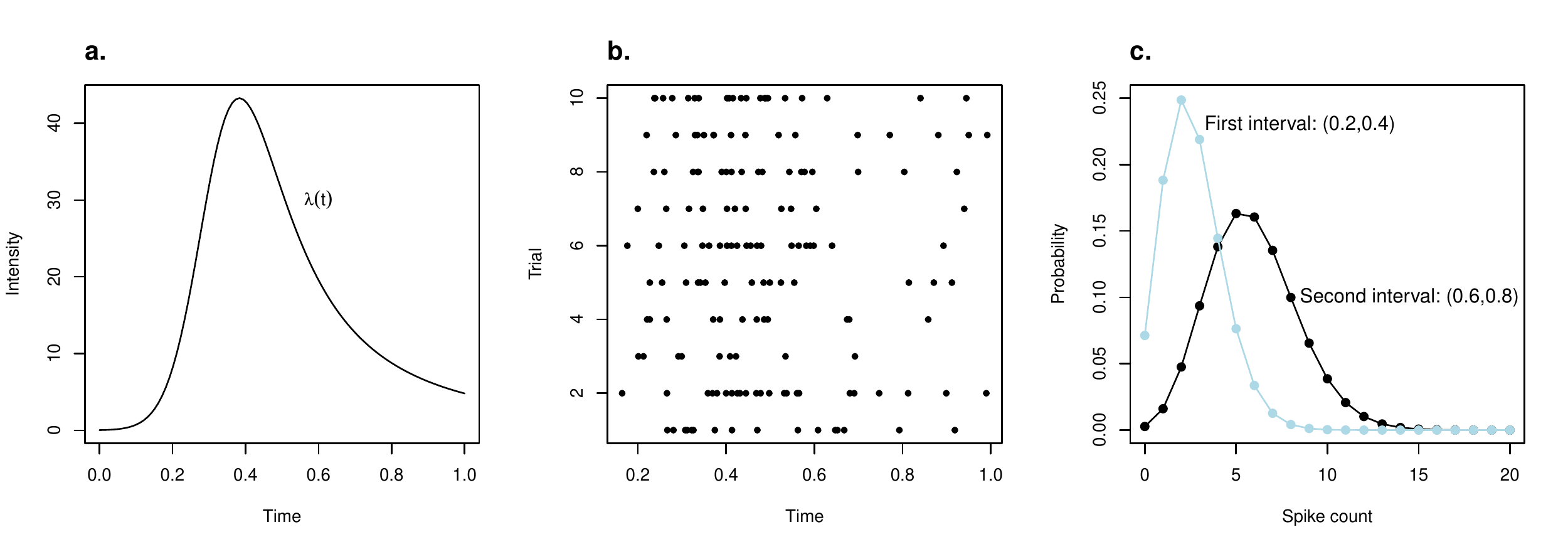} 
\par\end{centering}

\caption{A first example of a point process: the Inhomogeneous Poisson Process
(IPP) as a model for spike trains. \textbf{a. }The neuron is assumed
to respond to stimulation at a varying rate over time. The latent
rate is described by an intensity function, $\lambda(t)$ \textbf{b.
}Spikes are stochastic: here we simulated spike trains from an IPP
with intensity $\lambda(t)$. Different trials correspond to different
realisations. Note that a given spike train can be seen simply as
a set of points in $(0,1)$. \textbf{c. }The defining property of
the IPP is that spike counts in a given interval follow a Poisson
distribution. Here we show the probability of observing a certain
number of spikes in two different time intervals. \label{fig:Example-Temporal-IPP}}
\end{figure}

\par\end{center}

\begin{figure}
\begin{centering}
\subfloat[The latent intensity function $\lambda(x,y)$ controls how many points
will fall on average in a certain spatial area. Higher intensities
are in lighter shades of blue.]{\includegraphics[scale=0.5]{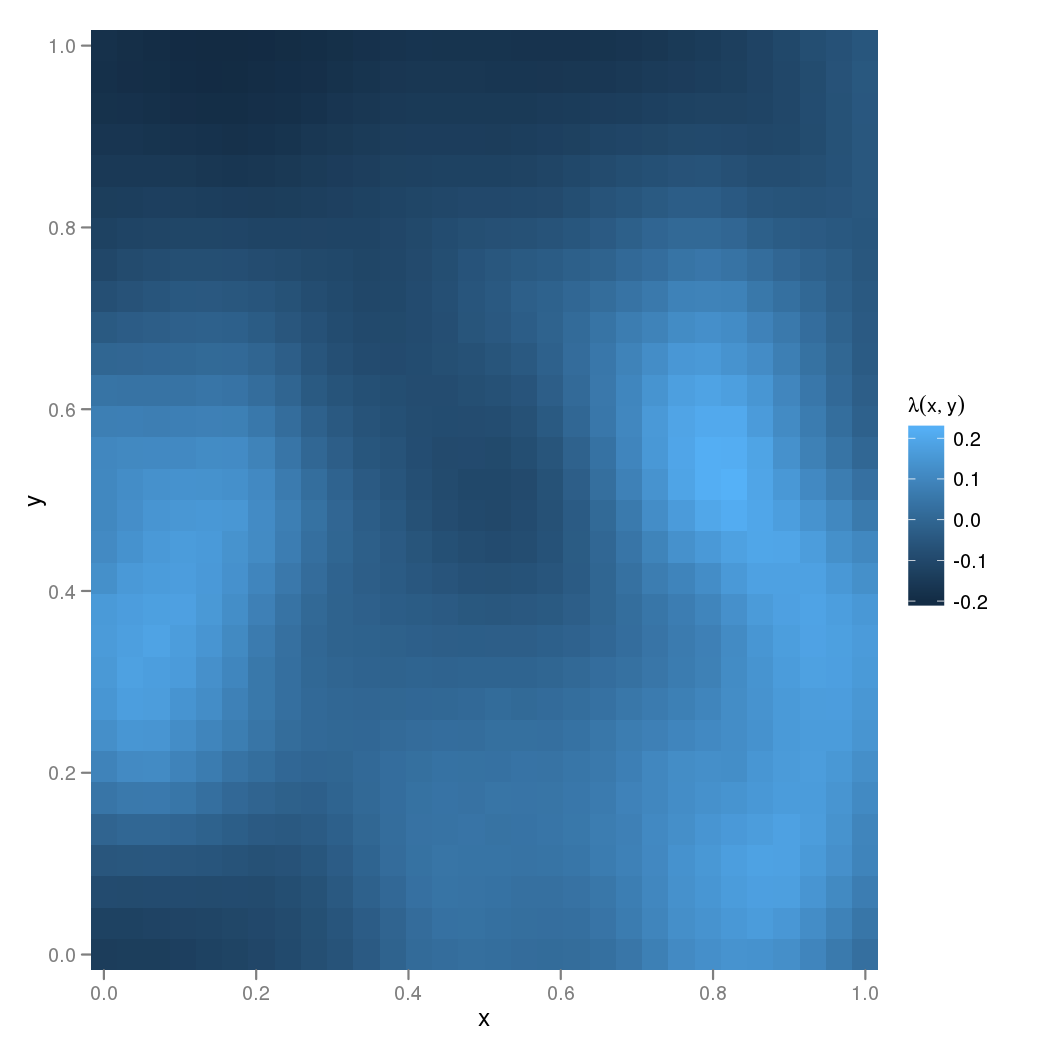}

}
\par\end{centering}

\begin{centering}
\subfloat[Four samples from an IPP with the intensity function shown in the
left-hand panel.]{\includegraphics[scale=0.5]{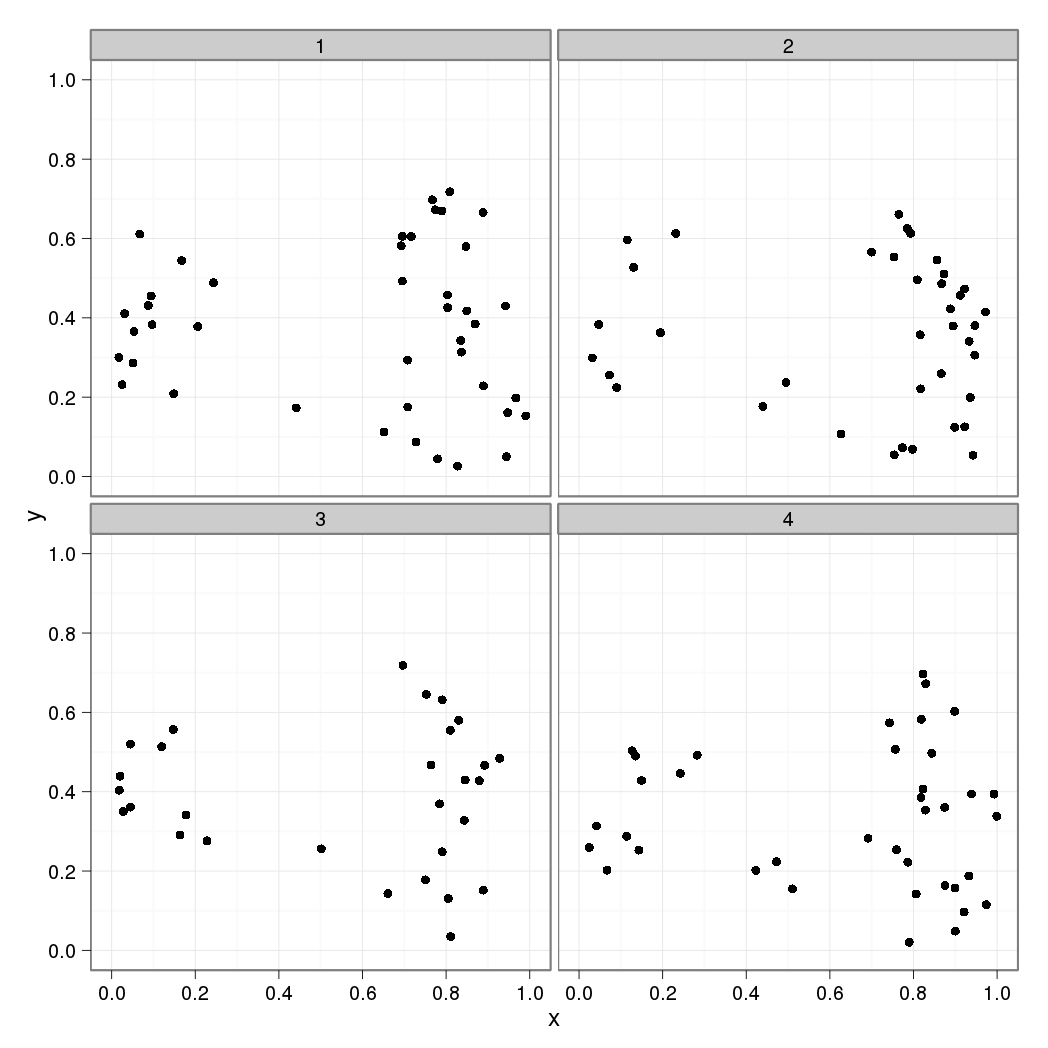}

}
\par\end{centering}

\caption{The spatial IPP is a mathematically straightforward extension to the
temporal IPP introduced in Figure \ref{fig:Example-Temporal-IPP}.
The main ingredient is a spatial intensity function $\lambda(x,y)$.
The IPP produces random point sets, as in the lower panel. When analysing
point data, the goal is usually to recover the latent intensity function
from such samples.\label{fig:Example-Spatial-IPP}}
\end{figure}

\subsection{The point of point processes}

Given a point set, the most natural question to ask is, generally,
``what latent intensity function could have generated the observed
pattern?'' Indeed, we argue that a lot of very specific research
questions are actually special cases of this general problem.

For mathematical convenience, we will from now on focus on the log-intensity
function $\eta(x,y)=\log\lambda\left(x,y\right)$. The reason this
is more convenient is that $\lambda\left(x,y\right)$ cannot be negative---and
we are not expecting a negative number of points (fixations)---whereas
$\eta\left(x,y\right)$, on the other hand, can take any value whatever,
from minus to plus infinity.

At this point we need to introduce the notion of \emph{spatial covariate},
which are directly analoguous to covariates in linear models. In statistical
parlance, the \emph{response }is what we are interested in predicting,
and \emph{covariates }is what we use to predict the response with.
In the case of point processes covariates are often spatial functions
too.

One of the classical questions in the study of overt attention is
the role of low-level cues in attracting gaze (i.e.\ visual saliency).
Among low-level cues, local contrast may play a prominent role, and
it is a classical finding that observers tend to be more interested
in high-contrast regions when viewing natural images, e.g.\ \citep{Rajashekar:FoveatedAnalysisImgFeatFixation}.

Imagine that our point set $\mathbf{F}=\left\{ \left(x_{1},y_{1}\right),\ldots,\left(x_{n},y_{n}\right)\right\} $
represents observed fixation locations on a certain image, and we
assume that these fixation locations were generated by an IPP with
log-intensity function $\eta\left(x,y\right)$. We suppose that the
value of $\eta\left(x,y\right)$ at location $x,y$ has something
to do with the local contrast $c(x,y)$ at the location. In other
words, the image contrast function $c(x,y)$ will enter as a \emph{covariate
}in our model. The simplest way to do so is to posit that $\eta\left(x,y\right)$
is a linear function of $c(x,y)$, i.e.:

\begin{equation}
\eta\left(x,y\right)=\beta_{c}\times c(x,y)+\beta_{0}\label{eq:simple-contrast-covariate}
\end{equation}

We have introduced two free parameters, $\beta_{c}$ and $\beta_{0}$,
that will need to be estimated from the data. $\beta_{c}$ is the
more informative of the two: for example, a positive value indicates
that high contrast is predictive of high intensity, and a nearly-null
value indicates that contrast is not related to intensity (or at least
not monotonically). We will return to this idea below when we consider
analysing the output of low-level saliency models.

Another example that will come up in our analysis is the well-documented
issue of the ``centrality bias'', whereby human observers in psychophysical
experiments in front of a centrally placed computer screen tend to
fixate central locations more often regardless of what they are shown
\citep{Tatler:CentralFixationBias}. Again this has an influence on
the intensity function that needs to be accounted for. One could postulate
another spatial (intrinsic) covariate, $d\left(x,y\right)$, representing
the distance to the centre of the display: e.g., $d(x,y)=\sqrt{x^{2}+y^{2}}$
assuming the centre is at $(0,0)$. As in Equation (\ref{eq:simple-contrast-covariate}),
we could write

\[
\eta\left(x,y\right)=\beta_{d}\times d(x,y)+\beta_{0}
\]

However, in a given image, both centrality bias and local contrast
will play a role, resulting in: 

\begin{equation}
\eta\left(x,y\right)=\beta_{d}\times d(x,y)+\beta_{c}\times c(x,y)+\beta_{0}\label{eq:example-additive-decomp}
\end{equation}

The relative contribution of each factor will be determined by the
relative values of $\beta_{d}$ and $\beta_{c}$. Such additive decompositions
will be central to our analysis, and we will cover them in much more
detail below.

\section{Case study: Analysis of low-level saliency models \label{sec:Case-study-1}}


If eye movement guidance is a relatively inflexible system which uses
local image cues as heuristics for finding interesting places in a
stimulus, then low-level image cues should be predictive of where
people look when they have nothing particular to do. This has been
investigated many times (see \citealp{Schuetz:EyeMovementsPerceptionReview}),
and there are now many datasets available of ``free-viewing'' eye
movements in natural images \citep{VanDerLinde:DOVES,Torralba:ContextualGuidanceEyeMovements}.
Here we use the dataset of \citet{Kienzle:CenterSurroundPatternsOptimalPredictors}
because the authors were particularly careful to eliminate a number
of potential biases (photographer's bias, among other things).

In \citet{Kienzle:CenterSurroundPatternsOptimalPredictors}, subjects
viewed photographs taken in a zoo in Southern Germany. Each image
appeared for a short, randomly varying duration of around 2 sec%
\footnote{The actual duration was sampled from a Gaussian distribution $\N\left(2,0.5^{2}\right)$
truncated at 1 sec. %
}. Subjects were instructed to ``look around the scene'', with no
particular goal given. The raw signal recorded from the eye-tracker
was processed to yield a set of saccades and fixations, and here we
focus only on the latter. We have already mentioned in the introduction
that such data are often analysed in terms of a patch classification
problem: can we tell between fixated and non-fixated image patches
based on their content? We now explain how to replicate the main features
of a such an analysis in terms of the point process framework.

\subsection{Understanding the role of covariates in determining fixated locations}

To be able to move beyond the basic statement that local image cues
somehow \emph{correlate }with fixation locations, it is important
that we clarify how covariates could enter into the latent intensity
function. There are many different ways in which this could happen,
with important consequences for the modelling. Our approach is to
build a model gradually, starting from simplistic assumptions and
introducing complexity as needed.

To begin with we imagine that local contrast is the only cue that
matters. A very unrealistic but drastically simple model assumes that
the more contrast there is in a region, the more subjects' attention
will be attracted to it. In our framework we could specify this model
as:

\[
\eta\left(x,y\right)=\beta_{0}+\beta_{1}c(x,y)
\]

However, surely other things besides contrast matters - what about
average luminance, for example? Couldn't brighter regions attract
gaze?

This would lead us to expand our model to include luminance as another
spatial covariate, so that the log-intensity function becomes:\\

\[
\eta\left(x,y\right)=\beta_{0}+\beta_{1}c(x,y)+\beta_{2}l(x,y)
\]

in which $l(x,y)$ stands for local luminance. But perhaps edges matter,
so why not include another covariate corresponding to the output of
a local edge detector $e(x,y)$? This results in:

\[
\eta\left(x,y\right)=\beta_{0}+\beta_{1}c(x,y)+\beta_{2}l(x,y)+\beta_{3}e(x,y)
\]

It is possible to go further down this path, and add as many covariates
as one sees fit (although with too many covariates, problems of variable
selection do arise, see \citealp{Hastie:ESL}), but to make our lives
simpler we can also rely on some prior work in the area and use pre-existing,
off-the-shelf \emph{image-based saliency models} \citep{FecteauMunoz:SalienceRelevancePriorityMap}.
Such models combine many local cues into one interest map, which saves
us from having to choose a set of covariates and then estimating their
relative weight (although see \citealp{Vincent:DoWeLookAtLights}
for work in a related direction). Here we focus on the perhaps most
well-known among these models, described in \citet{IttiKoch:ComputationalModellingVisualAttention}
and \citet{WaltherKoch:ModelingAttentionSalientProtoObjects}, although
many other interesting options are available (e.g., \citealp{BruceTsotsos:SaliencyAttentionVisualSearch},
\citealp{ZhaoKoch:LearningSaliencyMap}, or \citealp{Kienzle:CenterSurroundPatternsOptimalPredictors}).

\begin{figure}
\begin{centering}
\includegraphics[scale=0.45]{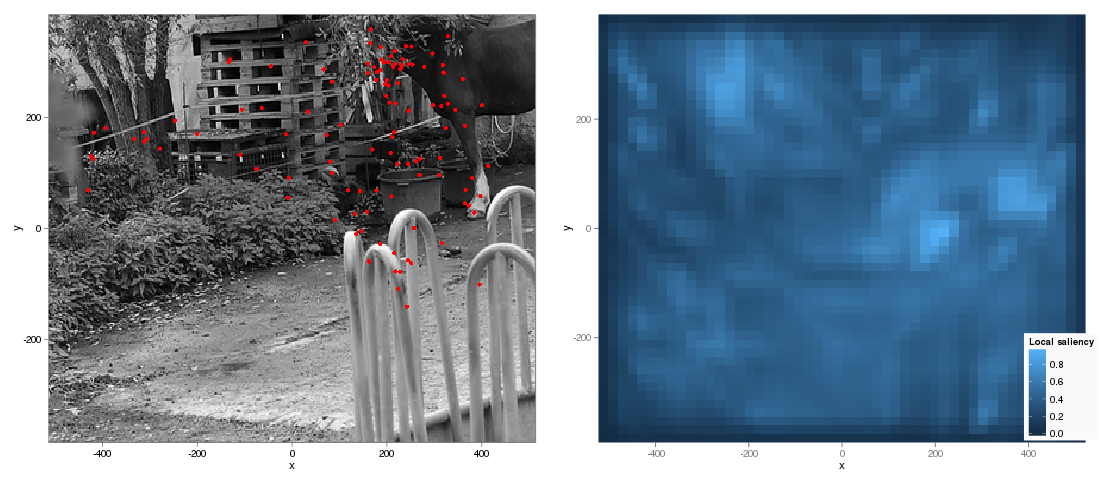} 
\par\end{centering}

\caption{An image from the dataset of \citet{Kienzle:CenterSurroundPatternsOptimalPredictors},
along with an ``interest map'' - local saliency computed according
to the Itti-Koch model \citep{IttiKoch:ComputationalModellingVisualAttention,WaltherKoch:ModelingAttentionSalientProtoObjects}.
Fixations made by the subjects are overlaid in red. How well does
the interest map characterise this fixation pattern? This question
is not easily answered by eye, but may be given a more precise meaning
in the context of spatial processes. \label{fig:IttiKochSaliencyKienzle}}
\end{figure}

The model computes several feature maps (orientation, contrast, etc.)
according to physiologically plausible mechanisms, and combines them
into one master map which aims to predict what the interesting features
in image $i$ are. For a given image $i$ we can obtain the interest
map $m_{i}\left(x,y\right)$ and use that as the unique covariate
in a point process:

\begin{equation}
\eta_{i}\left(x,y\right)=\alpha_{i}+\beta_{i}m_{i}(x,y)\label{eq:itti-saliency-simple}
\end{equation}

\begin{figure}
\begin{centering}
\includegraphics{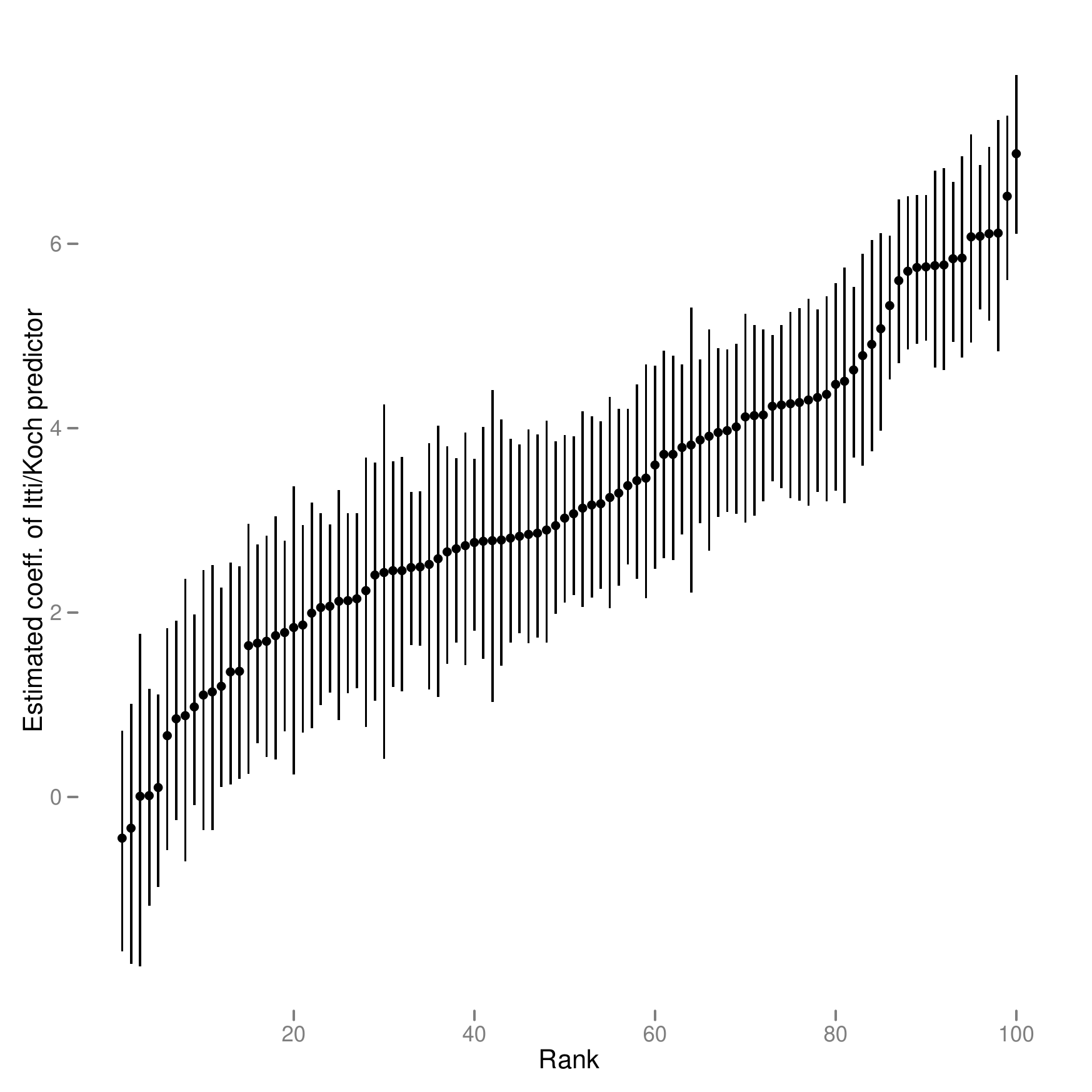} 
\par\end{centering}

\caption{Variability in the predictivity of the Itti-Koch model across images.
We estimate $\beta_{i}$ in Equation \ref{eq:itti-saliency-simple}
for 100 different images from the dataset of \citet{Kienzle:CenterSurroundPatternsOptimalPredictors}.
We plot the sorted mean-a-posteriori estimates along with a 95\% Bayesian
credible interval. The results show clearly that the ``interestingness''
given by low-level saliency is of variable value when predicting fixations:
for some images $\beta_{i}$ is very close to 0, which indicates that
there is no discernible association between low-level saliency and
fixation intensity in these images. In other images the association
is much stronger. \label{fig:Variability-IK-coeffs}}
\end{figure}

This last equation will be the starting point of our modelling. We
have changed the notation somewhat to reflect some of the adjustments
we need to make in order to learn anything from applying model to
data. To summarise: 
\begin{itemize}
\item $\eta_{i}(x,y)$ denotes the log-intensity function for image $i$,
which depends on the spatial covariate $m_{i}\left(x,y\right)$ that
corresponds to the interest map given by the low-level saliency of
\citet{IttiKoch:ComputationalModellingVisualAttention}. 
\item $\beta_{i}$ is an image-specific coefficient that measures to what
extent spatial intensity can be predicted from the interest map. $\beta_{i}=0$
means no relation, $\beta_{i}>0$ means that higher low-level saliency
is associated on average with more fixations, $\beta_{i}<0$ indicates
the opposite - people looked more often at low points of the interest
map. We make $\beta_{i}$ image-dependent because we anticipate that
how well the interest map predicts fixations depends on the image,
an assumption that is borne out, as we will see. 
\item $\alpha_{i}$ is an image specific intercept. It is required for technical
reasons but otherwise plays no important role in our analysis. 
\end{itemize}
We fitted the model given by Equation (\ref{eq:itti-saliency-simple})
to a dataset consisting of the fixations recorded in the first 100
images of the dataset of \citet[see Fig. ref{fig:IttiKochSaliencyKienzle}]{Kienzle:CenterSurroundPatternsOptimalPredictors}.
Computational methods are described in the appendix. We obtained a
set of posterior estimates for the $\beta_{i}$'s, of which a summary
is given in Figure \ref{fig:Variability-IK-coeffs}.

To make the coefficients shown on Figure \ref{fig:Variability-IK-coeffs}
more readily interpretable, we have scaled $m_{i}\left(x,y\right)$
so that in each image the most interesting points (according to the
Itti-Koch model) have value 1 and the least interesting 0. In terms
of the estimated coefficients $\beta_{i}$, this implies that the
intensity ratio between a maximally interesting region and a minimally
interesting region is equal to $e^{\beta_{i}}$: for example, a value
of $\beta_{i}$ of 1 indicates that in image $i$ on average a region
with high ``interestingness'' receives roughly 2.5 more fixations
than a region with very low ``interestingness''. At the opposite
end of the spectrum, in images in which the Itti-Koch model performs
very well, we have values of $\beta_{i}\approx6$, which implies a
ratio of 150 to 1 for the most interesting regions compared to the
least interesting.

\begin{figure}
\begin{centering}
\includegraphics[angle=90,scale=0.4]{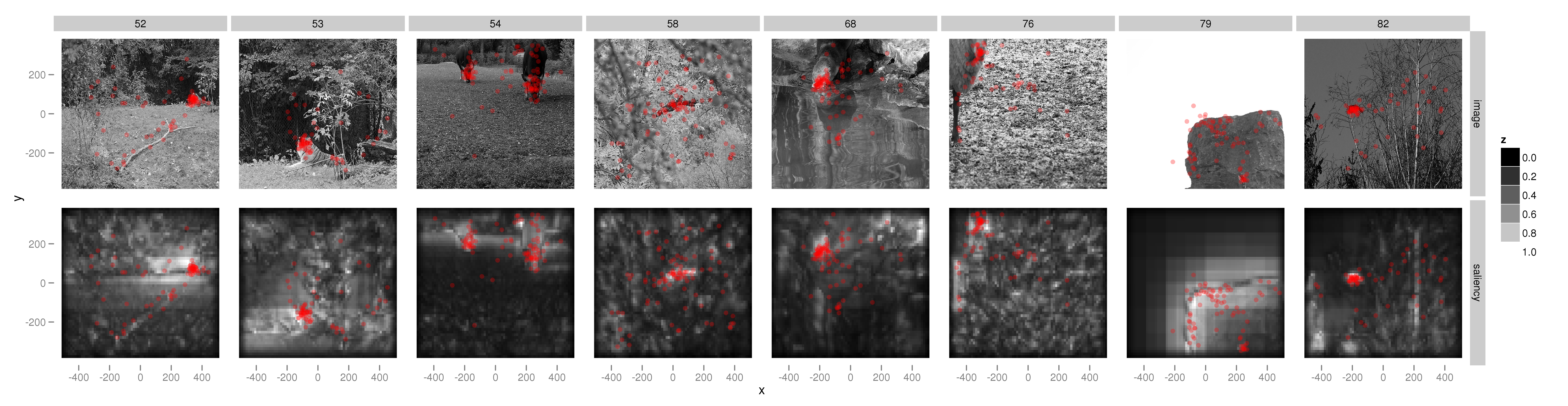} 
\par\end{centering}

\caption{Out of first 100 pictures in the Kienzle et al. dataset, we show the
8 ones in which Itti\&Koch interestingness has the strongest link
to fixation density (according to the value of $\beta_{i}$ in Equation
\ref{eq:itti-saliency-simple}). The I\&K interest map is displayed
below each image, and fixation locations are in red. \label{fig:Itti-best}}
\end{figure}

\begin{figure}
\begin{centering}
\includegraphics[angle=90,scale=0.4]{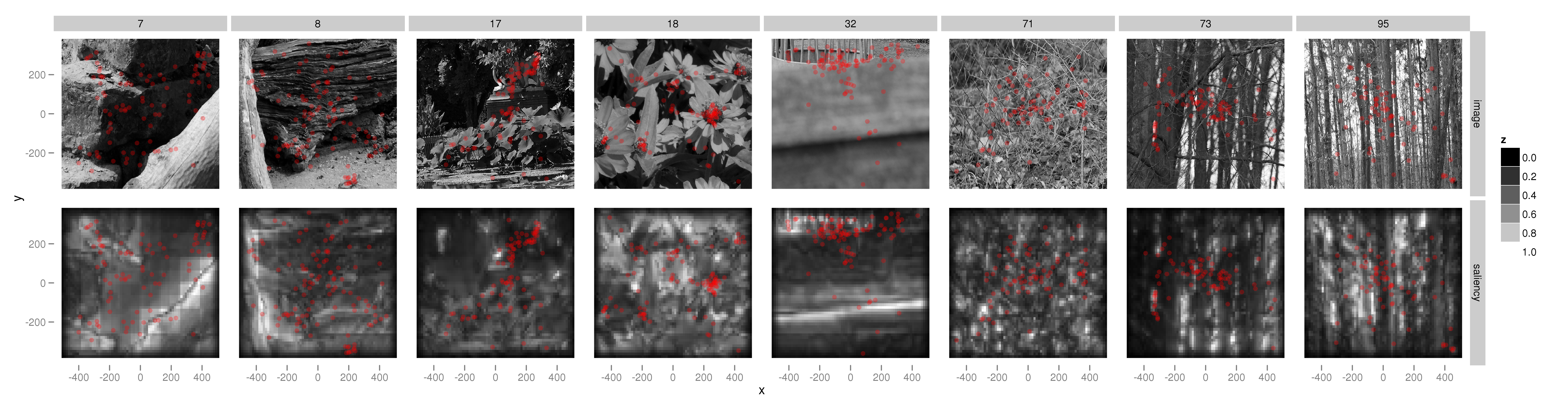} 
\par\end{centering}

\caption{Same as Figure \ref{fig:Itti-best} above, but with the 8 images with
lowest value for $\beta_{i}$ . \label{fig:Itti-worst}}
\end{figure}

It is instructive to compare the images in which the model does well%
\footnote{$\beta_{i}$ should not be interpreted as anything more than a rough
measure of performance. It has a relatively subtle potential flaw:
if the Itti-Koch map for an image happens by chance to match the typical
spatial bias, then $\beta_{i}$ will likely be estimated to be above
0. This flaw is corrected when a spatial bias term is introduced,
see Section \ref{sub:Including-a-spatial-bias}. %
}, to those in which it does poorly. On Figure \ref{fig:Itti-best}
we show the 8 images with highest $\beta_{i}$ value, and on Figure
\ref{fig:Itti-worst} the 8 images with lowest $\beta_{i}$, along
with the corresponding Itti-Koch interest maps. It is evident that,
while on certain images the model does extremely well, for example
when it manages to pick up the animal in the picture (see the lion
in images 52 and 53), in others it gets fooled by high-contrast edges
that subjects find highly uninteresting. Foliage and rock seem to
be particularly difficult, at least from the limited evidence available
here.

Given a larger annotated dataset, it would be possible to confirm
whether the model performs better for certain categories of images
than others. Although this is outside the scope of the current paper,
we would like to point out that the model in Equation (\ref{eq:itti-saliency-simple})
can be easily extended for that purpose: If we assume that images
are encoded as being either ``foliage'' or ``not foliage'', we
may then define a variable $\phi_{i}$ that is equal to 1 if image
$i$ is foliage and 0 if not. We may re-express the latent log-intensity
as:

\[
\eta_{i}\left(x,y\right)=\alpha_{i}+\left(\phi_{i}\gamma+\delta_{i}\right)m_{i}(x,y)
\]

which decomposes $\beta_{i}$ as the sum of an image-specific effect
($\delta_{i}$) and an effect of belonging to the foliage category
$\left(\gamma\right)$. Having $\gamma<0$ would indicate that pictures
of foliage are indeed more difficult on average%
\footnote{This may not necessarily be a intrinsic flaw of the model: it might
well be that in certain ``boring'' pictures, or pictures with very
many high-contrast edges, people will fixate just about anywhere,
so that even a perfect model---the \textquotedbl{}true\textquotedbl{}
causal model in the head of the observers---would perform relatively
badly. %
}. We take foliage here only as an illustration of the technique, as
it is certainly not the most useful categorical distinction one could
make (for a taxonomy of natural images, see \citealp{FeiFei:WhatDoWePercInAGlance},
and, e.g., \citealp{KasparKoenig:ViewingBehaviorImpactLowLevelImgProp}
for a discussion of image categories).

\begin{figure}
\begin{centering}
\includegraphics[scale=0.5]{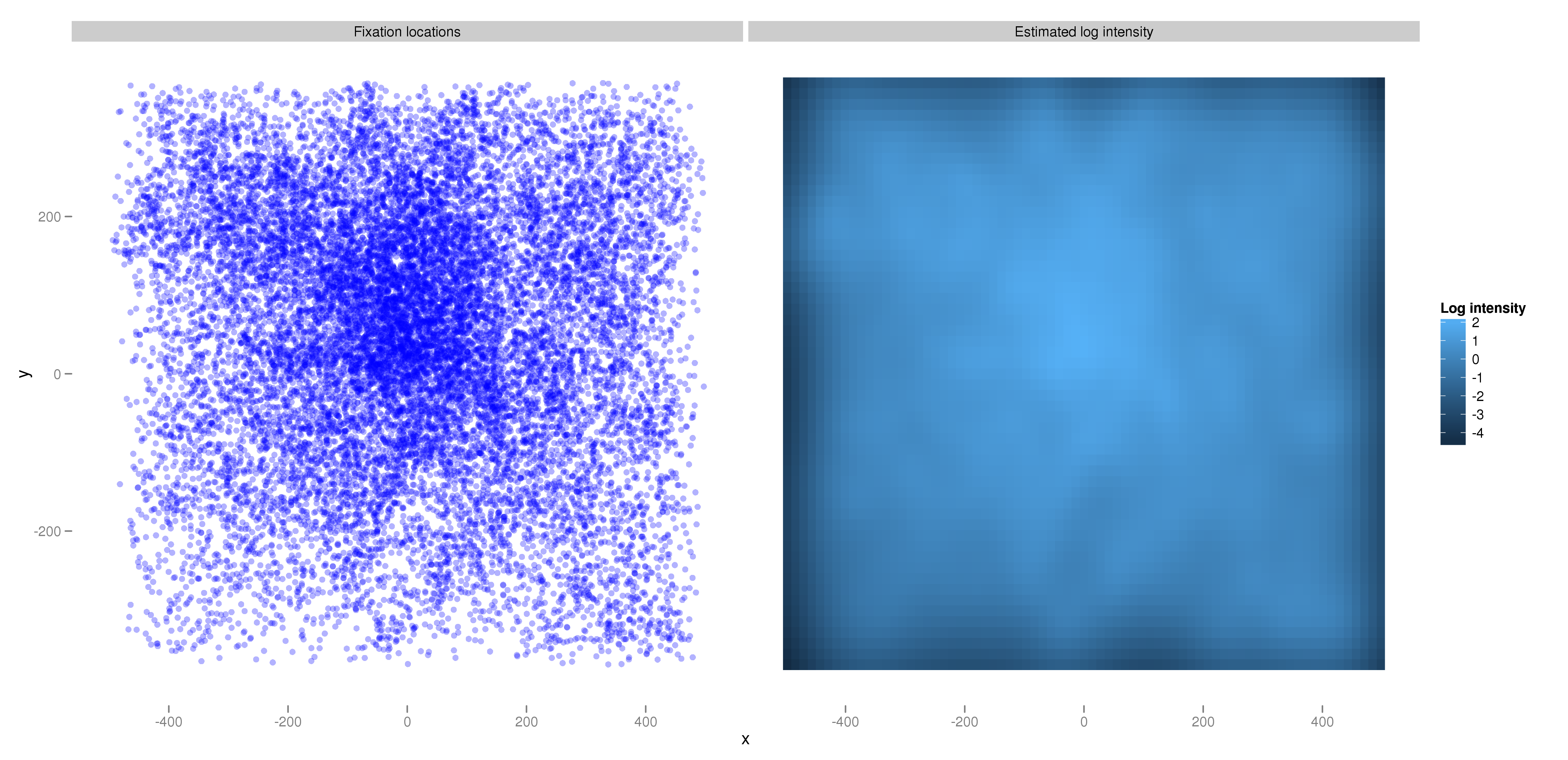} 
\par\end{centering}

\caption{The centrality bias. On the left panel, we plot every fixation recorded
in \citet{Kienzle:CenterSurroundPatternsOptimalPredictors}. On the
right, a non-parametric Bayesian estimate of the intensity function.
Central locations are much more likely to be fixated than peripheral
ones. \label{fig:Centrality-bias}}
\end{figure}

A related suggestion \citep{Torralba:ContextualGuidanceEyeMovements}
is to augment low-level saliency models with some higher-level concepts,
adding face detectors, text detectors, or horizon detectors. Within
the limits of our framework, a much easier way to improve predictions
is to take into account the \emph{centrality bias }\citep{TatlerVincent:BehavBiasesEyeGuidance},
i.e. the tendency for observers to fixate more often at the centre
of the image than around the periphery. One explanation for the centrality
bias is that it is essentially a side-effect of photographer's bias:
people are interested in the centre because the centre is where photographers
usually put the interesting things, unless they are particularly incompetent.
In \citet{Kienzle:CenterSurroundPatternsOptimalPredictors} photographic
incompetence was simulated by randomly clipping actual photographs
so that central locations were not more likely to be interesting than
peripheral ones. The centrality bias persists (see Fig. \ref{fig:Centrality-bias}),
which shows that central locations are preferred regardless of image
content (a point already made in \citealp{Tatler:CentralFixationBias}).
We can use this fact to make better predictions by making the required
modifications to the intensity function.

Before we can explain how to do that, we need to introduce a number
of additional concepts. A central theme in the proposed spatial point
process framework is to develop tools that help us to understand performance
of our models in detail. In the next section we introduce some relatively
user-friendly graphical tools for assessing fit. We will also show
how one can estimate an intensity function in a non-parametric way,
that is, without assuming that the intensity function has a specific
form. Nonparametric estimates are important in their own right for
visualisation (see for example the right-hand-side of Fig. \ref{fig:Centrality-bias}),
but also as a central element in more sophisticated analyses.

\subsection{Graphical model diagnostics\label{sub:Graphical-model-diagnostics}}

Once one has fitted a statistical model to data, one has to make sure
the fitted model is actually at least in rough agreement with the
data. 
A good fit alone is not the only thing we require of a model, because
fits can in some cases be arbitrarily good if enough free parameters
are introduced (see e.g., \citealp{BishopPRML}, ch. 3). But assessing
fit is an important step in model criticism \citep{GelmanHill:DataAnalysisUsingRegression},
which will let us diagnose model failures, and in many cases will
enable us to obtain a better understanding of the data itself. In
this section we will focus on informal, graphical diagnostics. More
advanced tools are described in \citet{Baddeley:ResidualAnalysisSpatialPP}.
\citet{Ehinger:ModelingSearchForPeopleIn900Scenes} use a similar
model-criticism approach in the context of saliency modelling.

Since a statistical model is in essence a recipe for how the data
are generated, the most obvious thing to do is to compare data simulated
from the model to the actual data we measured. In the analysis presented
above, the assumption is that the data come from a Poisson process
whose log-intensity is a linear function of Itti-Koch interestingness:

\begin{equation}
\eta_{i}\left(x,y\right)=\alpha_{i}+\beta_{i}m_{i}(x,y)\label{eq:itti-saliency-simple-1}
\end{equation}

\begin{figure}
\begin{centering}
\includegraphics[scale=0.4]{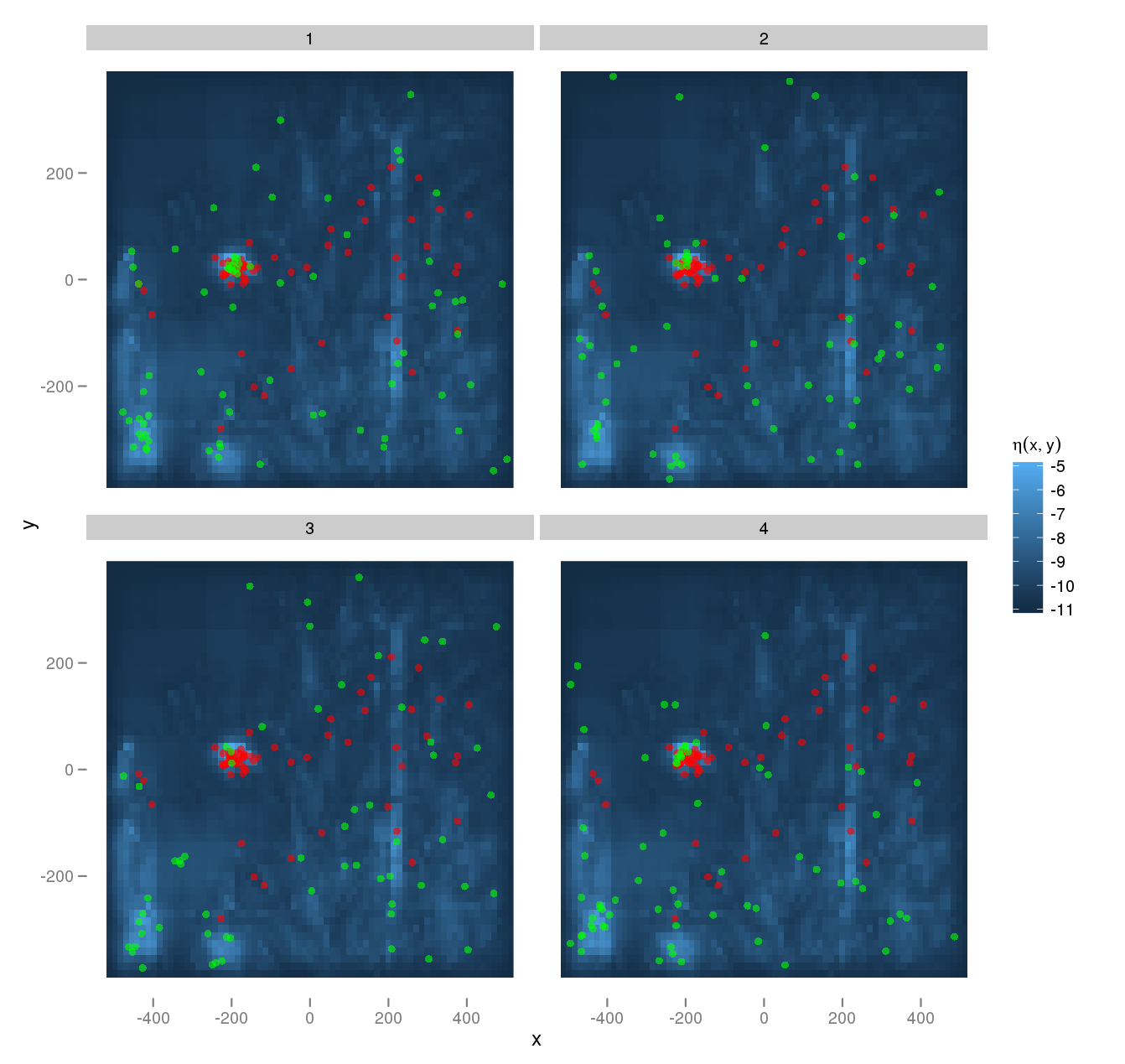} 
\par\end{centering}

\caption{Simulating from a fitted point process model. The fixations on image
82 (rightmost in Figure \ref{fig:Itti-best}) were fitted with the
model given by Equation (\ref{eq:itti-saliency-simple}), resulting
in an estimated log-intensity $\eta(x,y)$ which is plotted as a heatmap
in the background of each panel. In red we plot the actual fixation
locations (the same in every of the four panels), and in green simulations
from the fitted model, four different realizations, one in each panel.
\label{fig:simulations-fitted-model}}
\end{figure}

For a given image, we have estimated values $\hat{\alpha}_{i},\hat{\beta}_{i}$
(mean a posterior estimate). A natural thing to do is to ask what
data simulated from a model with those parameters look like%
\footnote{Simulation from an IPP can be done using the ``thinning'' algorithm
of \citet{LewisShedler:SimulationNonhomogenPP}, which is a form of
rejection sampling.%
}. In Figure \ref{fig:simulations-fitted-model}, we take the image
with the maximum estimated value for $\beta_{i}$ and compare the
actual recorded fixation locations to four different simulations from
an IPP with the fitted intensity function.

\begin{figure}
\begin{centering}
\includegraphics[scale=0.4]{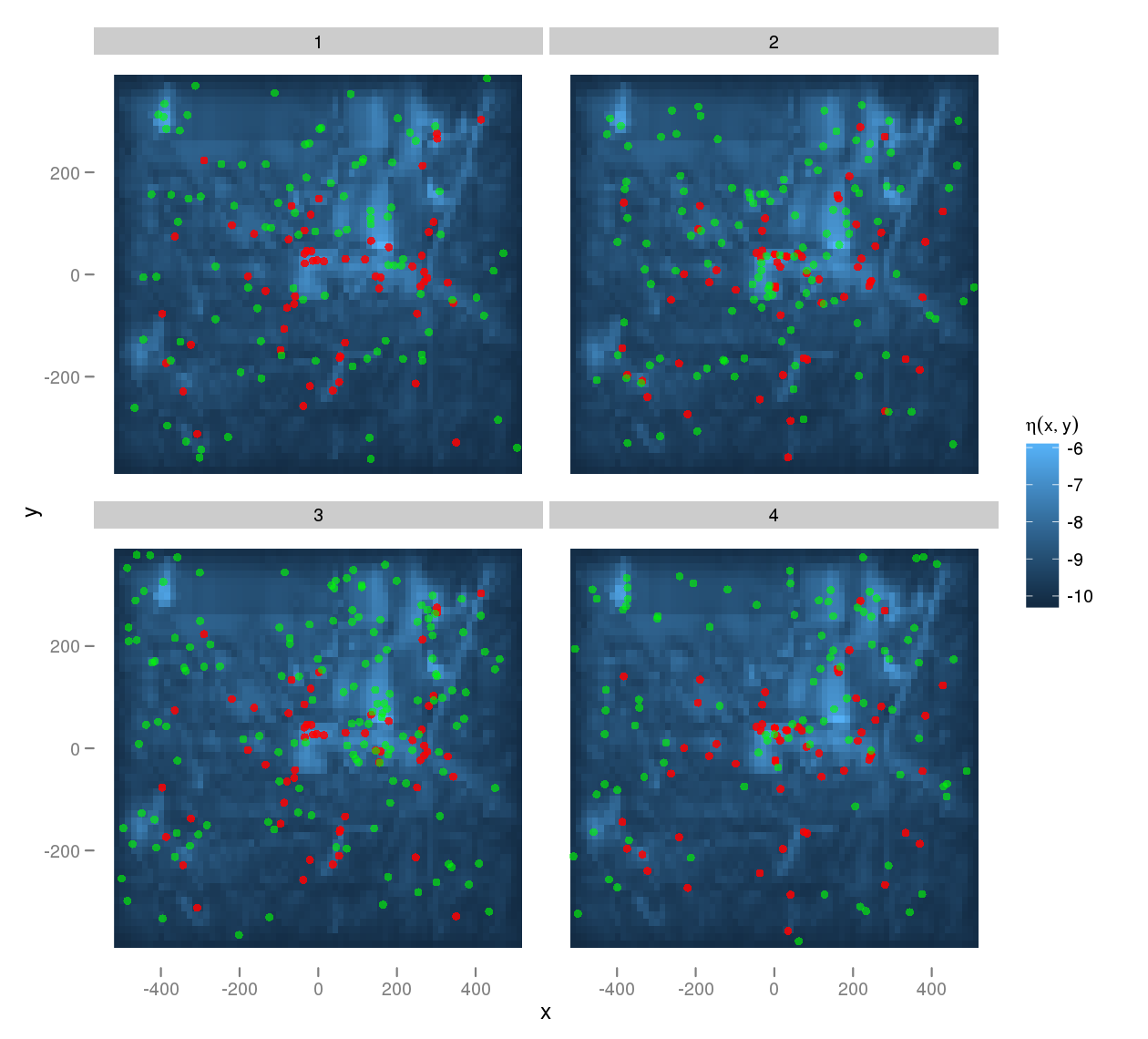}~\includegraphics[scale=0.4]{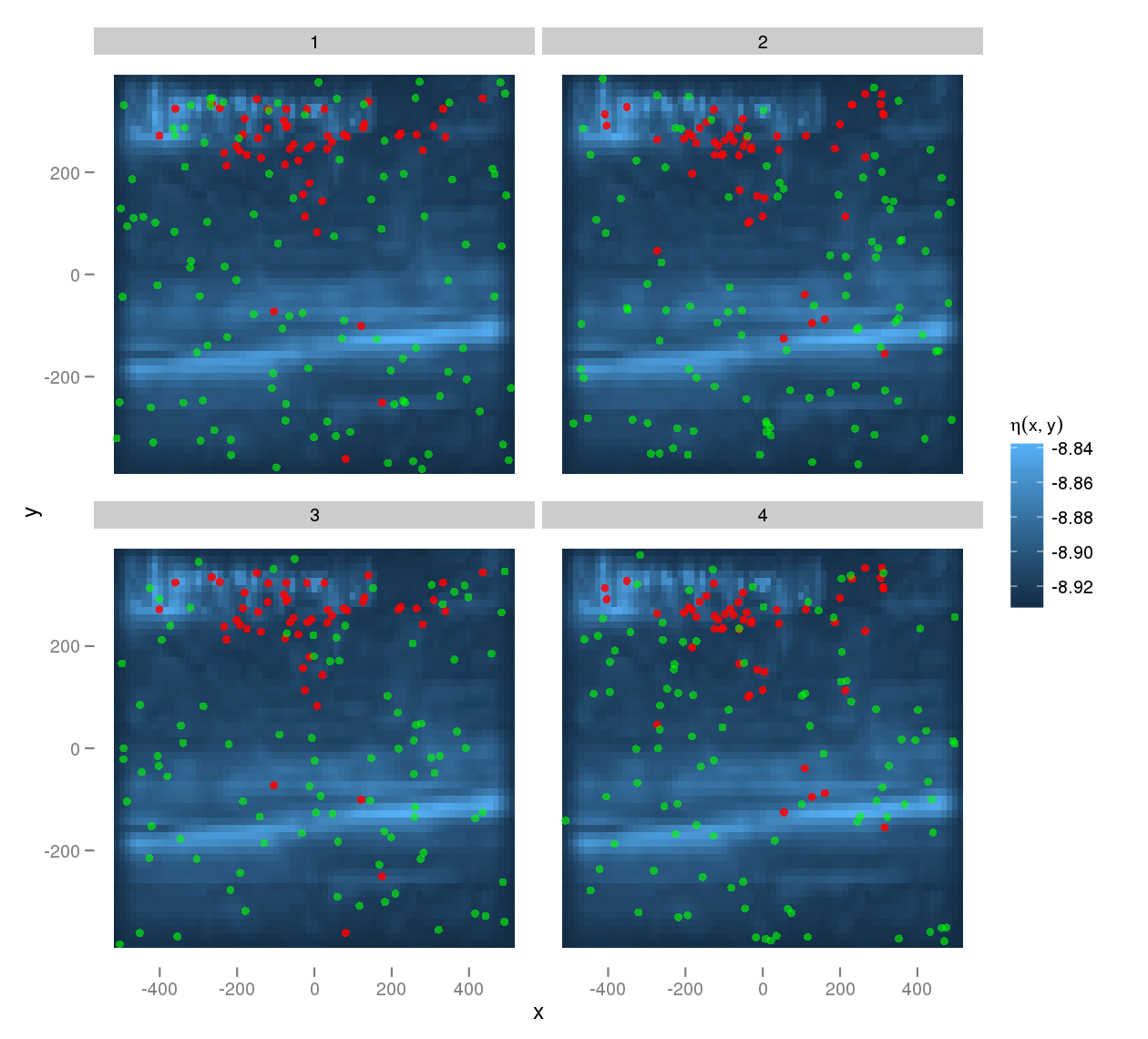} 
\par\end{centering}

\caption{Same as in Figure \ref{fig:simulations-fitted-model}, with the fixations
measured on image 45 (left) and 32 (right) of the dataset. The agreement
between data and simulations is of distinctly poorer quality than
in image 82.\label{fig:Simulations-fitted-less-good}}
\end{figure}

What is immediately visible from the simulations is that, while the
real data present one strong cluster that also appears in the simulations,
the simulations have a higher proportion of points outside of the
cluster, in areas far from any actual fixated locations. Despite these
problems, the fit seems to be quite good compared to other examples
from the dataset: Figure \ref{fig:Simulations-fitted-less-good} shows
two other examples, image 45, which has a median $\beta$ value of
about 4, and image 32, which had a $\beta$ value of about 0. In the
case of image 32, since there is essentially no relationship between
the interestingness values and fixation locations, the best possible
intensity function of the form given by Equation (\ref{eq:itti-saliency-simple})
is one with $\beta=0$, a uniform intensity function.

It is also quite useful to inspect some of the \emph{marginal }distributions.
By marginal distributions we mean point distributions that we obtain
by merging data from different conditions. In Figure \ref{fig:marginal-fixation-locations-simple},
we plot the fixation locations across all images in the dataset. In
the lower panel we compare it to simulations from the fitted model,
in which we generated fixation locations from the fitted model for
each image so as to simulate an entire dataset. This brings to light
a failure of the model that would not be obvious from looking at individual
images: based on Itti-Koch interestingness alone we would predict
a distribution of fixation locations that is almost uniform, whereas
the actual distribution exhibits a central bias, as well as a bias
for the upper part of the screen.

\begin{center}
\begin{figure}
\begin{centering}
\includegraphics[scale=0.6]{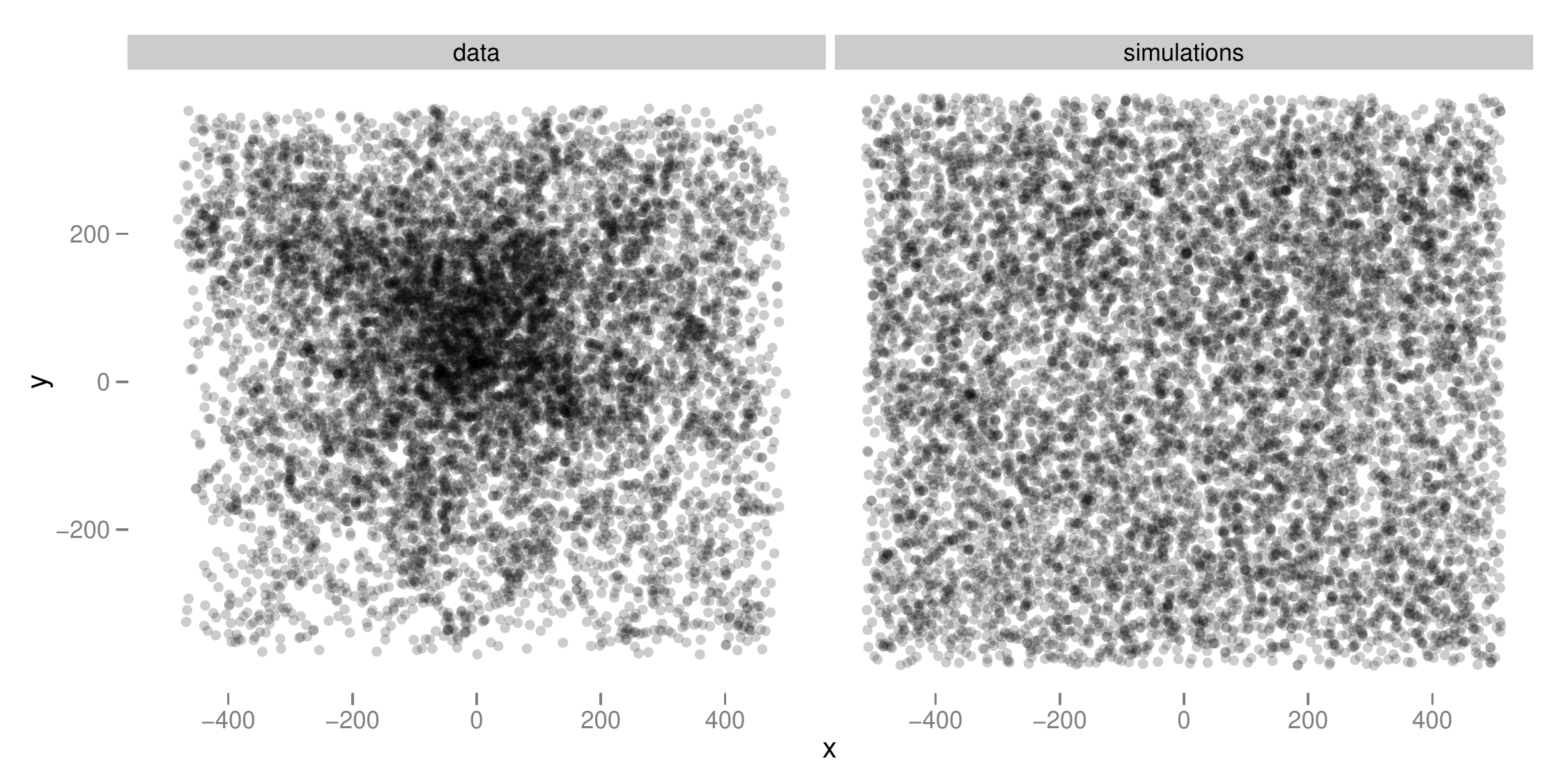} 
\par\end{centering}

\caption{Comparing marginal fixation locations. On the left panel, we plot
each fixation location in images 1 to 100 (each dot corresponds to
one fixation). On the right panel, we plot simulated fixation locations
from the fitted model corresponding to Equation (\ref{eq:itti-saliency-simple}).
A strong spatial bias is visible in the data, not captured at all
by the model.\label{fig:marginal-fixation-locations-simple}}
\end{figure}

\par\end{center}

Overall, the model derived from fitting Equation (\ref{eq:itti-saliency-simple})
seems rather inadequate, and we need to account at least for what
seems to be some prior bias favouring certain locations. Explaining
how to do so requires a short detour through the topic of non-parametric
inference, to which we turn next.

\subsection{Inferring the intensity function non-parametrically\label{sub:Inferring-the-intensity-function-nonparam}}

Consider the data in Figure \ref{fig:Centrality-bias}: to get a sense
of how much observers prefer central locations relative to peripheral
ones, we could define a central region \textbf{$\mathcal{A}$}, count
how many fixations fall in it, compared to how many fixations fall
outside. From the theoretical point of view, what we are doing is
directly related to estimating the intensity function: the expected
number of fixations in $\mathcal{A}$ is after all $\int_{\mathcal{A}}\lambda\left(x,y\right)\mbox{d}x\mbox{d}y$,
the integral of the intensity function over $\mathcal{A}$. Seen the
other way, counting how many sample points are in $\mathcal{A}$ is
a way of estimating the integral of the intensity over $\mathcal{A}$.

Modern statistical modelling emphasizes non-parametric estimation.
If one is trying to infer the form of an unknown function $f(x)$,
one should not assume that $f(x)$ has a certain parametric form unless
there is very good reason for this choice (interpretability, actual
prior knowledge or computational feasibility). Assuming a parametric
form means assuming for example that $f(x)$ is linear, or quadratic:
in general it means assuming that $f(x)$ can be written as $f(x)=\phi(x;\beta)$,
where $\beta$ is a finite set of unknown parameters, and $\phi\left(x;\beta\right)$
is a family of functions over $x$ parameterised by $\beta$. Nonparametric
methods make much weaker assumptions, usually assuming only that $f$
is smooth at some spatial scale.

We noted above that estimating the integral of the intensity function
over a spatial region could be done by counting the number of points
the region contains. Assume we want to estimate the intensity $\lambda(x,y)$
at a certain point $x_{0},y_{0}$. We have a realisation $S$ of the
point process (for example a set of fixation locations). If we assume
that $\lambda(x,y)$ is spatially smooth, it implies that $\lambda(x,y)$
varies slowly around $x_{0},y_{0}$, so that we may consider it roughly
constant in a small region around $x_{0},y_{0}$, for instance in
a circle of radius $r$ around $(x_{0},y_{0})$. Call this region
$\mathcal{C}_{r}$ - the integral of the intensity function over $\mathcal{C}_{r}$
is related to the intensity at $(x_{0}$,$y_{0}$) in the following
way:

\[
\int_{\mathcal{C}_{r}}\lambda(x,y)\mbox{d}x\mbox{d}y\approx\int_{\mathcal{C}_{r}}\lambda\left(x_{0},y_{0}\right)\mbox{d}x\mbox{d}y=\lambda\left(x_{0},y_{0}\right)\times\int_{\mathcal{C}_{r}}\mbox{d}x\mbox{d}y
\]

$\int_{\mathcal{C}_{r}}\mbox{d}x\mbox{d}y$ is just the area of circle
$\mathcal{C}_{r}$, equal to $\pi r$. Since we can estimate $\int_{\mathcal{C}_{r}}\lambda(x,y)\mbox{d}x\mbox{d}y$
via the number of points in $\mathcal{C}_{r}$, it follows that we
can estimate $\lambda(x_{0},y_{0})$ via:

\[
\hat{\lambda}\left(x_{0},y_{0}\right)=\frac{\left|S\cap\mathcal{C}_{r}\right|}{\pi r}
\]

$\left|S\cap\mathcal{C}_{r}\right|$ is the cardinal of the intersection
of the point set $S$ and the circle $\mathcal{C}_{r}$ (note that
they are both sets), shorthand for ``number of points in $S$ that
are also in $\mathcal{C}_{r}$''.

\begin{figure}
\begin{centering}
\includegraphics[scale=0.45]{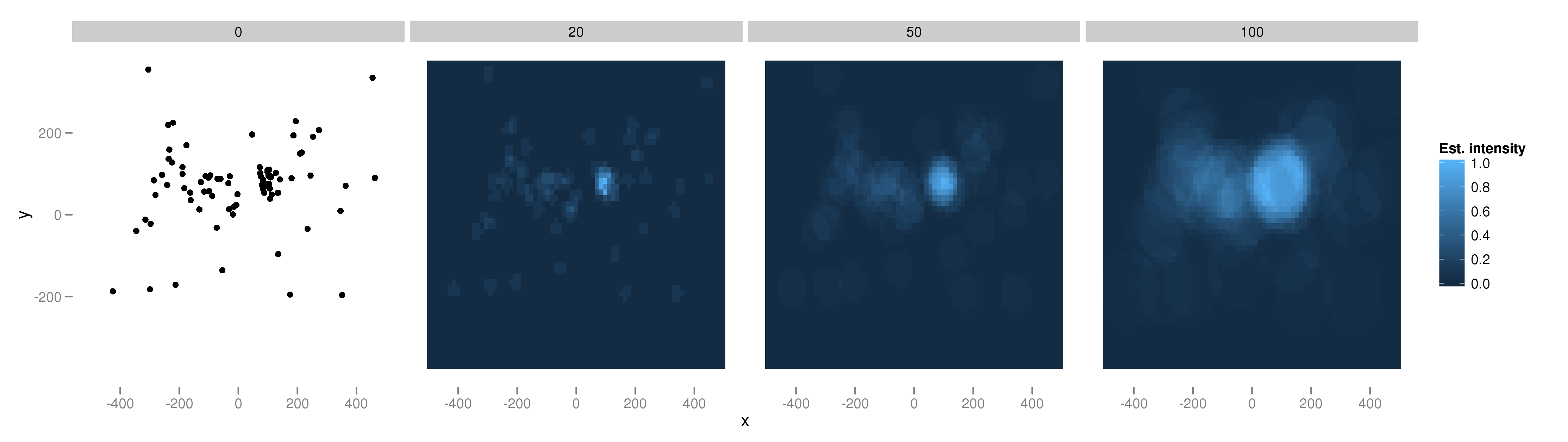} 
\par\end{centering}

\caption{Nonparametric estimation of the intensity function using a moving
window. The data are shown on the leftmost panel. The intensity function
at $(x,y)$ is estimated by counting how many points are within a
radius $r$ of $(x,y)$. We show the results for $r=20,50,100$. Note
that with $r\rightarrow0$ we get back the raw data. For easier visualization,
we have scaled the intensity values such that the maximum intensity
is 1 in each panel. \label{fig:Nonparametric-estimation-classical}}
\end{figure}

What we did for $(x_{0},y_{0})$ remains true for all other points,
so that a valid strategy for estimating $\lambda(x,y)$ at any point
is to count how many points in $S$ are in the circle of radius $r$
around the location. The main underlying assumption is that $\lambda\left(x,y\right)$
is roughly constant over a region of radius $r$. This method will
be familiar to some readers in the context of non-parametric density
estimation, and indeed it is almost identical%
\footnote{Most often, instead of using a circular window, a Gaussian kernel
will be used.%
}. It is a perfectly valid strategy, detailed in \citet{Diggle:StatisticalAnalysisSpatialPointPatterns},
and its only major shortcoming is that the amount of smoothness (represented
by $r$) one arbitrarily uses in the analysis may change the results
quite dramatically (see Figure \ref{fig:Nonparametric-estimation-classical}).
Although it is possible to also estimate $r$ from the data, in practice
this may be difficult (see \citealp{Illian:StatAnalysisSpatPointPatterns},
Section 3.3).

\begin{figure}
\begin{centering}
\includegraphics[scale=0.5]{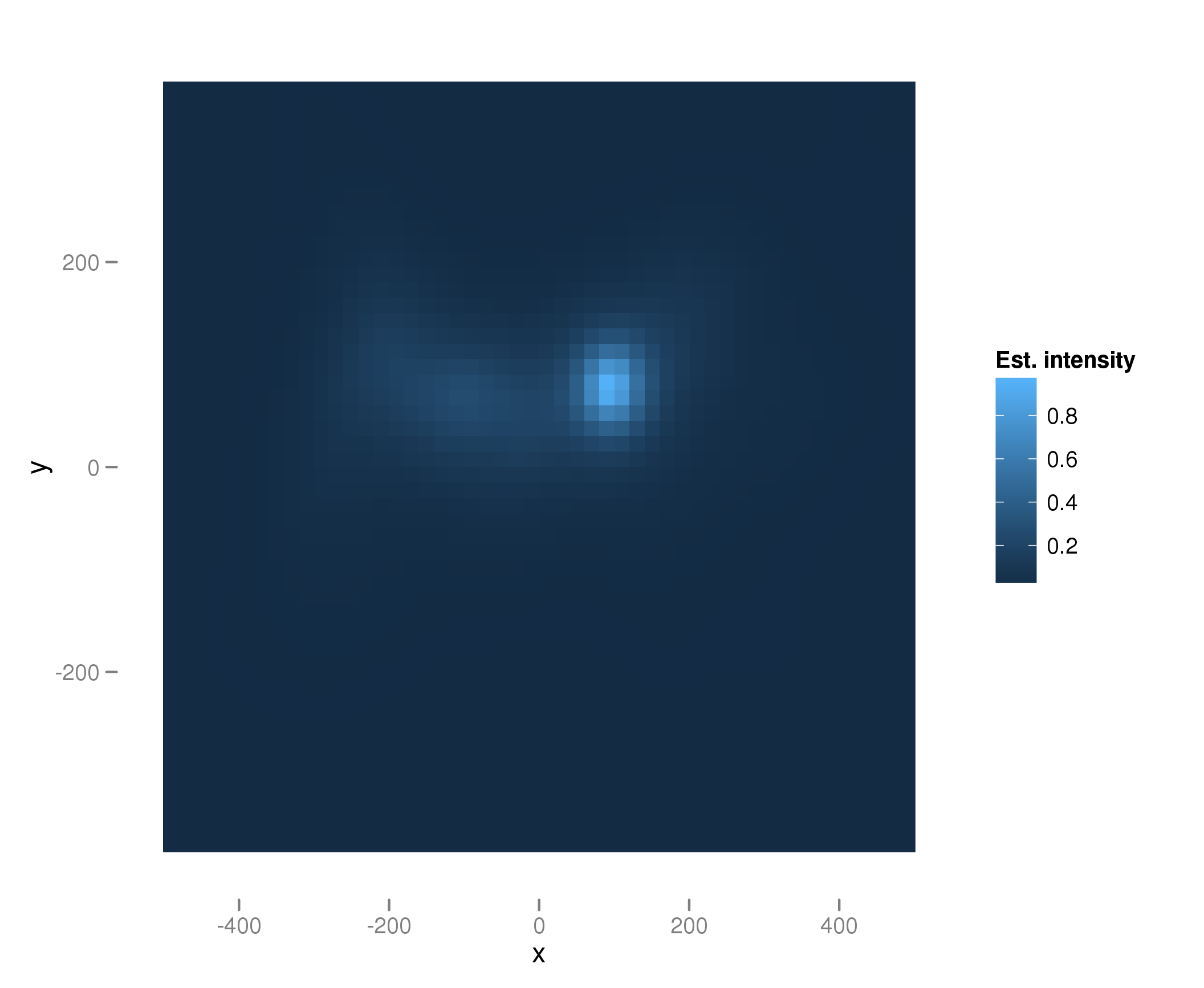} 
\par\end{centering}

\caption{Nonparametric Bayesian estimation of the intensity function. We use
the same data as in Figure \ref{fig:Nonparametric-estimation-classical}.
Inference is done by placing a Gaussian process prior on the log-intensity
function, which enforces smoothness. Hyperparameters are integrated
over. See text and appendix \ref{sub:Intro-to-GPs} for details. }
\end{figure}

There is a Bayesian alternative: put a prior distribution on the intensity
$\lambda$ and base the inference on the posterior distribution of
$\lambda(x,y)$ given the data, with

\[
p(\lambda|S)\propto p(S|\lambda)p(\lambda)
\]

as usual. We can use the posterior expectation of $\lambda(x,y)$
as an estimator (the posterior expectation is the mean value of $\lambda(x,y)$
given the data), and the posterior quantiles give confidence intervals%
\footnote{For technical reasons Bayesian inference is easier when done on the
log-intensity function $\eta(x,y)$, rather than on the intensity
function, so we actually use the posterior mean and quantiles of $\eta(x,y)$
rather than that of $\lambda(x,y)$.%
}. The general principles of Bayesian statistics will not be explained
here, the reader may refer to \citet{Kruschke:DoingBayesianAnalysis}
or any other of the many excellent textbooks on Bayesian statistics
for an introduction.

To be more precise, the method proceeds by writing down the very generic
model:

\[
\log\lambda\left(x,y\right)=f(x,y)+\beta_{0}
\]

and effectively forces $f(x,y)$ to be a relatively smooth function,
using a Gaussian Process prior. Exactly how this is achieved is explained
in Appendix \ref{sub:Intro-to-GPs}, but roughly, Gaussian Processes
let one define a probability distribution over functions such that
smooth functions are much more likely than non-smooth functions. The
exact spatial scale over which the function is smooth is unknown but
can be averaged over.

To estimate the intensity function of one individual point process,
there is little cause to prefer the Bayesian estimate over the classical
non-parametric estimate we described earlier. As we will see however,
using a prior that favours smooth functions becomes invaluable when
one considers \emph{multiple point processes} with shared elements.

\subsection{Including a spatial bias, and looking at predictions for new images\label{sub:Including-a-spatial-bias}}

We have established that models built from interest maps do not fit
the data very well, and we have hypothesized that one possible cause
might be the presence of a spatial bias. Certain locations might be
fixated despite having relatively uninteresting contents. A small
modification to our model offers a solution: we can hypothesize that
all latent intensity functions share a common component. In equation
form:

\begin{equation}
\eta_{i}\left(x,y\right)=\alpha_{i}+\beta_{i}m_{i}(x,y)+g(x,y)\label{eq:itti-koch-spatial-bias}
\end{equation}

\begin{figure}
\begin{centering}
\includegraphics[scale=0.6]{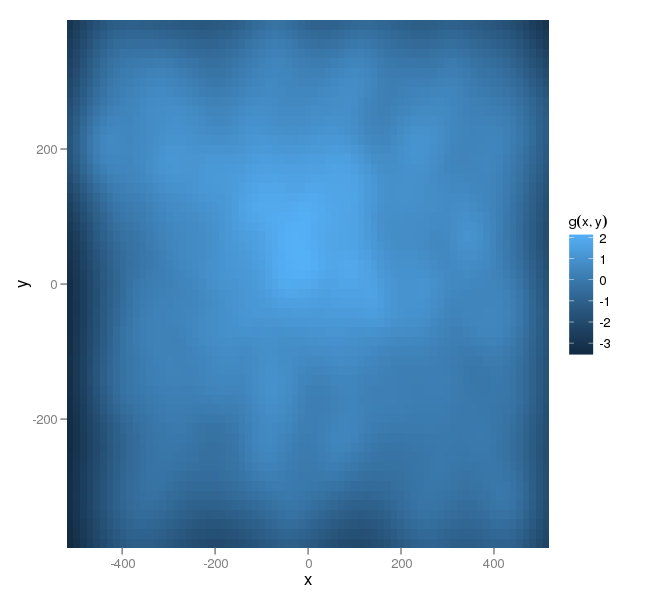} 
\par\end{centering}

\caption{Estimated spatial bias $g(x,y)$ (Eq. \ref{eq:itti-koch-spatial-bias}).\label{fig:itti-spatial-bias} }
\end{figure}

As in the previous section, we do not force $g(x,y)$ to take a specific
form, but only assume smoothness. Again, we use the first 100 images
of the dataset to estimate the parameters. The estimated spatial bias
is shown on Figure \ref{fig:itti-spatial-bias}. It features the centrality
bias and the preference for locations above the midline that were
already visible in the diagnostics plot of Section \ref{sub:Graphical-model-diagnostics}
(Fig. \ref{fig:marginal-fixation-locations-simple}).

From visual inspection alone it appears clear that including a spatial
bias is necessary, and that the model with spatial bias offers a significant
improvement over the one that does not. However, things are not always
as clear-cut, and one cannot necessarily argue from a better fit that
one has a better model. There are many techniques for statistical
model comparison, but given sufficient data the best is arguably to
compare the predictive performance of the different models in the
set (see \citealp{Pitt:TowardMethodSelectCompModelsCog}, and \citealp{Robert:TheBayesianChoice},
ch. 7 for overviews of model comparison techniques). In our case we
could imagine two distinct prediction scenarios: 
\begin{enumerate}
\item For each image, one is given, say, 80\% of the recorded fixations,
and must predict the remaining 20\%. 
\item One is given all fixation locations in the first $n$ images, and
must predict fixations locations in the next $k$. 
\end{enumerate}
To use low-level saliency maps in engineering applications \citep{Itti:AutomaticFoveationVideoCompression},
what is needed is a model that predicts fixation locations on arbitrary
images---i.e. the model needs to be good at the second prediction
scenario outlined above. The model is tuned on recorded fixations
on a set of training images, but to be useful it must make sensible
predictions for images outside the original training set.

From the statistical point of view, there is a crucial difference
between prediction problems (1) and (2) above. Problem (1) is easy:
to make predictions for the remaining fixations on image $i$, estimate
$\beta_{i}$ and $\alpha_{i}$ from the available data, and predict
based on the estimated values (or using the posterior predictive distribution).
Problem (2) is much more difficult: for a new image $j$ we have no
information about the values of $\beta_{j}$ or $\alpha_{j}$. In
other words, in a new image the interest map could be very good or
worthless, and we have no way of knowing that in advance.

We do however have information about what values $\beta_{j}$ and
$\alpha_{j}$ are likely to take from the estimated values for the
images in the training set. If in the training set nearly all values
$\beta_{1},\beta_{2},\ldots,\beta_{n}$ were above 0, it is unlikely
that $\beta_{j}$ will be negative. We can represent our uncertainty
about $\beta_{j}$ with a probability distribution, and this probability
distribution may be estimated from the estimated values for $\beta_{1},\beta_{2},\ldots,\beta_{n}$.
We could, for example, compute their mean and standard deviation,
and assume that $\beta_{j}$ is Gaussian distributed with this particular
mean and standard deviation%
\footnote{There is a cleaner way of doing that, using multilevel\slash{}random
effects modelling \citep{GelmanHill:DataAnalysisUsingRegression},
but a discussion of these techniques would take us outside the scope
of this work.%
}. Another way, which we adopt here, is to use a kernel density estimator
so as not to impose a Gaussian shape on the distribution.

As a technical aside: for the purpose of prediction the intercept
$\alpha_{j}$ can be ignored, as its role is to modulate the intensity
function globally, and it has no effect on where fixations happen,
simply on \emph{how many }fixations are predicted. Essentially, since
we are interested in fixation locations, and not in how many fixations
we get for a given image, we can safely ignore $\alpha_{i}.$ A more
mathematical argument is given in Appendix \ref{sub:Conditioning-on-n}.

Thus how to predict? We know how to predict fixation locations \emph{given
}a certain value of $\beta_{j}$, as we saw earlier in Section \ref{sub:Graphical-model-diagnostics}.
Since $\beta_{j}$ is unknown we need to average over our uncertainty.
A recipe for generating predictions is to sample a value for $\beta_{j}$
from $p(\beta_{j})$, and conditional on that value, sample fixation
locations. Please refer to Figure \ref{fig:Explanation-Predictions}
for an illustration.

\begin{center}
\begin{figure}
\begin{centering}
\includegraphics[scale=0.7]{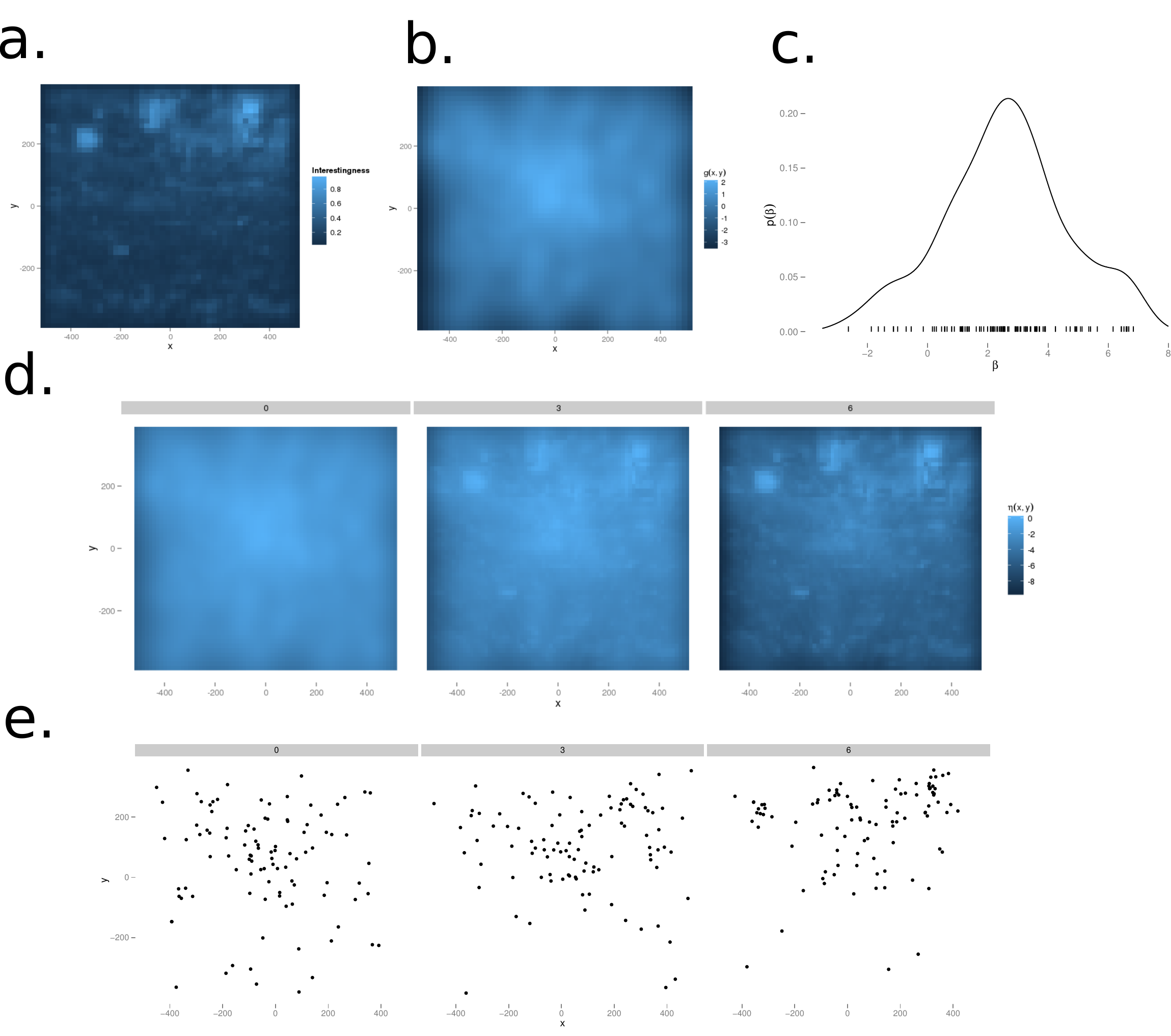} 
\par\end{centering}

\caption{Predictions for a novel image. The latent intensity function in Equation
\ref{eq:itti-koch-spatial-bias} has two important components: the
interest map $m(x,y)$, shown here in panel \textbf{(a)} for image
103, and a general spatial component $g(x,y)$, shown in \textbf{(b)}.
Image 103 does not belong to the training set, and the value of $\beta_{103}$
is therefore unknown: we do not know if $m(x,y)$ will be a strong
predictor or not, and must therefore take this uncertainty into account.
Uncertainty is represented by the distribution the $\beta$ coefficient
takes over images, and we can estimate this distribution from the
estimated values from the training set. In \textbf{(c)} we show those
values as dashes, along with a kernel density estimate. Conditional
on a given value for $\beta_{103}$, our predictions come from a point
process with log-intensity function given by $\beta_{103}m_{103}\left(x,y\right)+g\left(x,y\right)$:
in \textbf{(d)}, we show the intensity function for $\beta_{103}=0,3,6$.
In \textbf{(e)}, we show simulations from the corresponding point
processes (conditional on $n=100$ fixations, see \ref{sub:Conditioning-on-n}).
In general the strategy for simulating from the predictive distribution
will be to sample a value of $\beta$ from $p(\beta)$, and sample
from the corresponding point process as is done here. \label{fig:Explanation-Predictions}}
\end{figure}

\par\end{center}

In Figure \ref{fig:Predicting-marginal-fixations} we compare predictions
for marginal fixation locations (over all images), with and without
a spatial bias term. We simulated fixations from the predictive distribution
for images 101 to 200. We plot only one simulation, since all simulations
yield for all intents and purposes the same result: without a spatial
bias term, we replicate the problem seen in Figure \ref{fig:marginal-fixation-locations-simple}.
We predict fixations distributed more or less uniformly over the monitor.
Including a spatial bias term solves the problem.

\begin{center}
\begin{figure}
\begin{centering}
\includegraphics[scale=0.6]{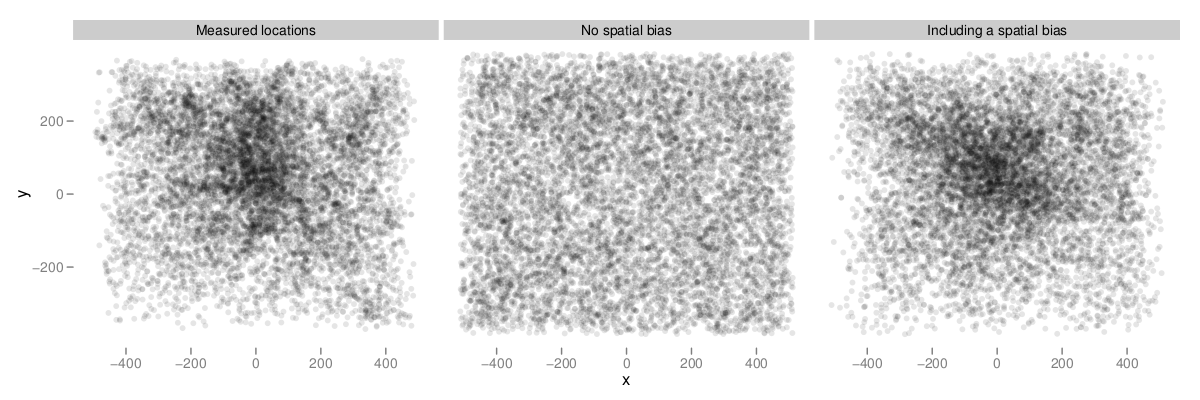} 
\par\end{centering}

\caption{Predicting marginal fixation locations. In the first panel we plot
the location of all fixations for images 101 to 200. In the second
panel we plot simulated fixation locations for the same images from
the naive model of Equation \ref{eq:itti-saliency-simple}. In the
second panel we plot simulated fixation locations for the same images
from the model of Equation \ref{fig:itti-spatial-bias}, which includes
a spatial bias. Note that these are predicted fixation locations for
entirely new images, and not a fit. Including a spatial bias improves
predictions enormously. \label{fig:Predicting-marginal-fixations}}
\end{figure}

\par\end{center}

What about predictions for individual images? Typically in vision
science we are attempting to predict a one-dimensional quantity: for
example, we might have a probability distribution for somebody's contrast
threshold. If this probability distribution has high variance, our
predictions for any \emph{individual} trial or the average of a number
of trials are by necessity imprecise. In the one-dimensional case
it is easy to visualise the degree of certainty by plotting the distribution
function, or providing a confidence interval. In a point process context,
we do not deal with one-dimensional quantities: if the goal is to
predict where 100 fixations on image $j$ might fall, we are dealing
with a 200 dimensional space---100 points times 2 spatial dimensions.
A maximally confident prediction would be represented by a probability
distribution that says that all points will be at a single location.
A minimally confident prediction would be represented by the uniform
distribution over the space of possible fixations, saying that all
possible configurations are equally likely. Thus the question that
needs to be addressed is, where do the predictions we can make from
the Itti-Koch model fall along this axis?

\begin{figure}
\begin{centering}
\includegraphics[scale=0.4]{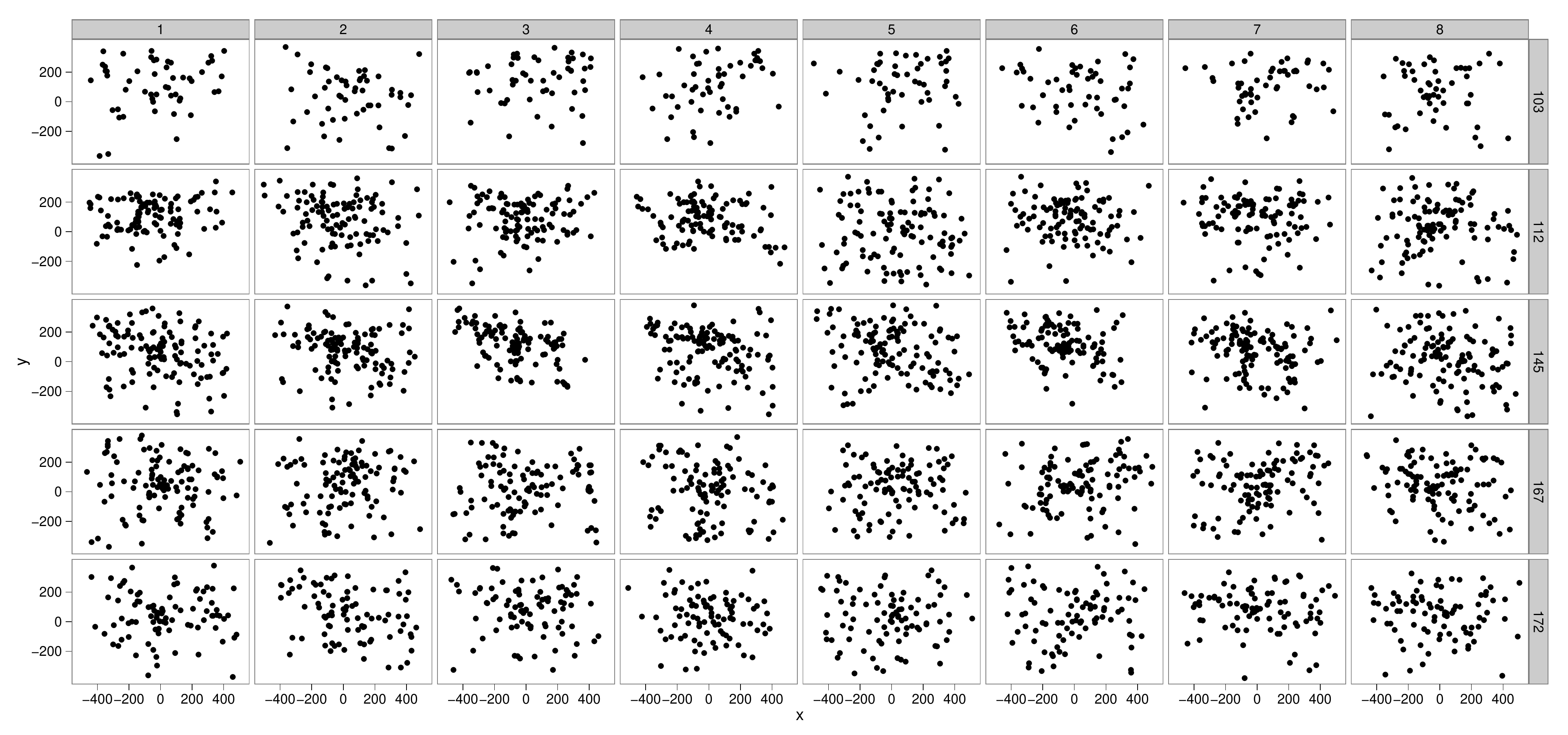} 
\par\end{centering}

\caption{Samples from the predictive distributions for the model including
spatial bias. We picked five images at random, and generated 8 samples
from the predictive distribution from each, using the technique outlined
in Figure \ref{fig:Explanation-Predictions}. Each row corresponds
to one image, with samples along the columns. This lets us visualise
the uncertainty in the predictive distributions, see text.\label{fig:Samples-predictive}}
\end{figure}

It is impossible to provide a probability distribution, or to report
confidence intervals. A way to visualise the amount of uncertainty
we have is by drawing samples from the predictive probability distribution,
to see if the samples vary a lot. Each sample is a set of a 100 points:
if we notice for example that over 10 samples all the points in each
sample systematically cluster at a certain location, it indicates
that our predictive distribution is rather specific. If we see at
lot of variability across samples, it is not. This mode of visualisation
is better adapted to a computer screen than to be printed on paper,
but for five examples we show eight samples in Figure \ref{fig:Samples-predictive}.

To better understand the level of uncertainty involved, imagine that
the objective is to perform (lossy) image compression. We picked this
example because saliency models are sometimes advocated in the compression
context \citep{Itti:AutomaticFoveationVideoCompression}. Lossy image
compression works by discarding information and hoping people will
not notice. The promise of image-based saliency models is that if
we can predict what part of an image people find interesting, we can
get away with discarding more information where people will not look.
Let us simplify the problem and assume that either we compress an
area or we do not. The goal is to find the largest possible section
of the image we can compress, under the constraint that if a 100 fixations
are made in the image, less than 5 fall in the compressed area (with
high probability). If the predictive distribution is uniform, we can
afford to compress less than 5\% of the area of the image. A full
formalisation of the problem for other distributions is rather complicated,
and would carry us outside the scope of this introduction, but looking
at the examples of Figure \ref{fig:Samples-predictive} it is not
hard to see qualitatively that for most images, the best area we can
find will be larger than 5\% but still rather small: in the predictive
distributions, points have a tendency of falling in most places except
around the borders of the screen.

The reason we see such imprecision in the predictive distributions
is essentially because we have to hedge our bets: since the value
of $\beta$ varies substantially from one image to another, our predictions
are vague by necessity. In most cases, models are evaluated in terms
of average performance (for example, average AUC performance over
the dataset). The above results suggest that looking just at average
performance is insufficient. A model that is consistenly mediocre
may for certain applications be preferable than a model that is occasionally
excellent but sometimes terrible. If we cannot tell in advance when
the latter model does well, our predictions about fixation locations
may be extremely imprecise.

\subsection{Dealing with non-stationarities: dependency on the first fixation\label{sub:Dealing-with-non-stationarities:}}

One very legitimate concern with the type of models we have used so
far is the independence assumption embedded into the IPP: all fixations
are considered independent of each other. Since successive fixations
tend to occur close to one another, we know that the independence
assumption is at best a rough approximation to the truth. There are
many examples of models in psychology that rather optimistically assume
independence and thus neglect nonstationarities: when fitting psychometric
functions for example, one conveniently ignores sequential dependencies,
learning, etc. but see \citet{Fruend:InfPsychFunNonstatBehav} or
Schönfelder and Wichmann (in press). One may argue, however, that
models assuming independence are typically simpler, and therefore
less likely to overfit, and that the presence of dependencies effectively
acts as an additional (zero-mean) noise source that does not bias
the results. This latter assumption requires to be explicitly checked,
however. In this section we focus specifically on a source of dependency
that could bias the results (dependency on the first fixation), and
show that (a) our models can amended to take this dependency into
account, (b) that the dependency indeed exists, although (c) results
are not affected in a major way. We conclude with a discussion of
dependent point processes and some pointers to the relevant literature.

\begin{figure}
\begin{centering}
\includegraphics[scale=0.8]{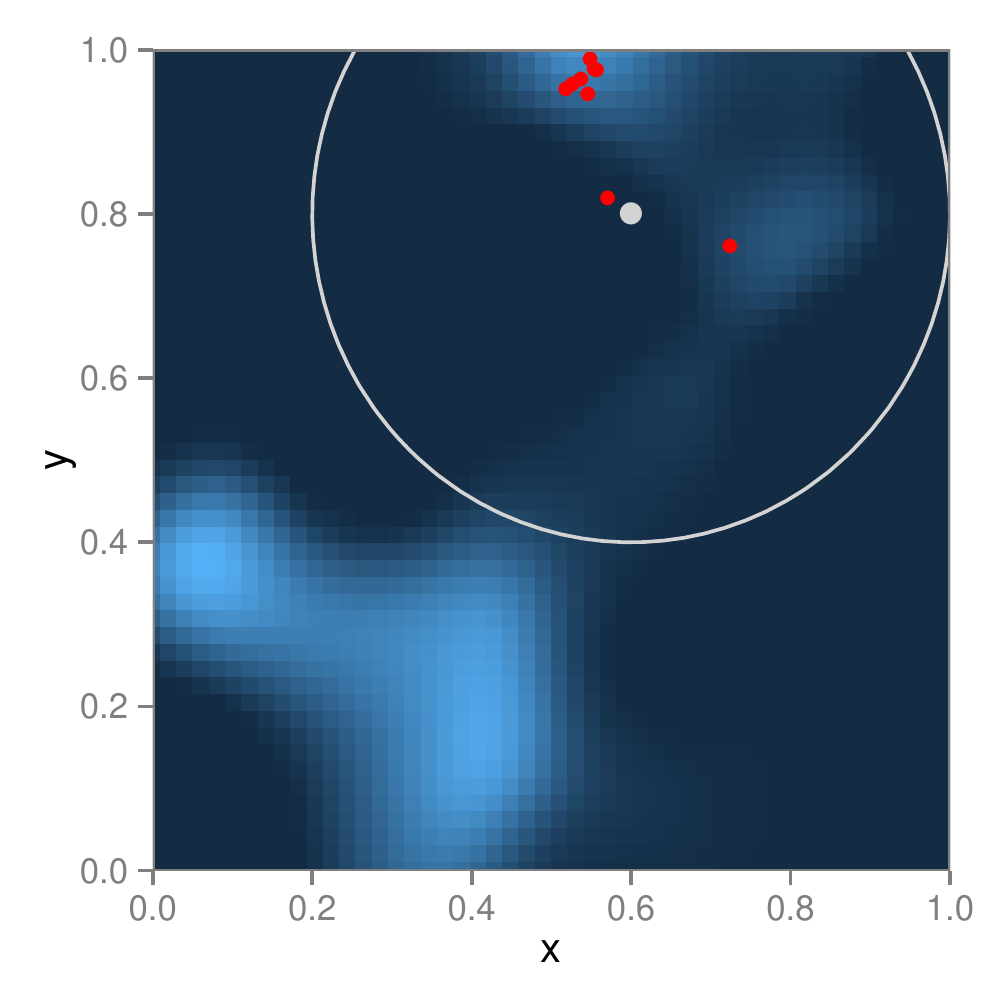} 
\par\end{centering}

\caption{Dependence on the initial fixation location: a potential source of
bias in estimation? We show here a random saliency map, and suppose
that fixation locations depend on saliency but are constrained by
how far away they can move from the original fixation location. The
original fixation is in grey, the circle shows a constraint region
and the other points are random fixations. The area at the bottom
half of the picture is left unexplored not because it is not salient
but because it is too far away from the original fixation location.
Such dependencies on the initial location may lead to an underestimation
of the role of saliency. We describe in the text a method for overcoming
the problem. \label{fig:dependence-on-initial-fixation}}
\end{figure}

In the experiment of \citet{Kienzle:CenterSurroundPatternsOptimalPredictors},
subjects only had a limited amount of time to look at the pictures:
generally less than 4 seconds. This does not always allow enough time
to explore the whole picture, so that subjects may have only explored
a part of the picture limited to an certain area around the initial
fixation. As explained on Figure \ref{fig:dependence-on-initial-fixation},
such dependence may cause us to underestimate the predictive value
of a saliency map. Supposing that fixations are indeed driven by the
saliency map, there might be highly salient regions that go unexplored
because they are too far away from the initial fixation. In a model
such as we have used so far, this problem would lead to under-estimating
the $\beta$ coefficients.

As with spatial bias, there is again a fairly straightforward solution:
we add an additional spatial covariate, representing distance to the
original fixation. The log-intensity functions are now modelled as:

\begin{equation}
\eta_{ij}\left(x,y\right)=\alpha_{ij}+\beta_{i}m_{i}(x,y)+\gamma d_{ij}(x,y)+\nu c(x,y)\label{eq:initial-fixation-model}
\end{equation}

Introducing this new spatial covariate requires some changes. Since
each subject started at a different location, we have one point process
per subject and per image, and therefore the log-intensity functions
$\eta_{ij}$ are now indexed both by image $i$ and subject $j$ (and
we introduce the corresponding intercepts $\alpha_{ij}$). The covariate
$d_{ij}\left(x,y\right)$ corresponds to the Euclidean distance to
the initial fixation, i.e. if the initial fixation of subject $j$
on image $i$ was at $x=10,y=50$, $d_{ij}\left(x,y\right)=\sqrt{\left(x-10\right)^{2}+\left(y-50\right)^{2}}$.
The coefficient $\gamma$ controls the effect of the distance to the
initial fixation: a negative value of $\gamma$ means that intensity
decreases away from the initial location, or in other words that fixations
tend to stay close to the initial location. For the sake of computational
simplicity, we have replaced the non-parametric spatial bias term
$g(x,y)$ with a linear term $\nu c(x,y)$ representing an effect
of the distance to the center ($c(x,y)=\sqrt{x^{2}+y{}^{2}}$). Coefficient
$\nu$ plays a role similar to $\gamma$: a negative value for $\nu$
indicates the presence of a centrality bias. We have scaled $c$ and
$d_{ij}$ so that a distance of 1 corresponds to the width of the
screen. In this analysis we exclude the initial fixations from the
set of fixations, they are used only as covariates. The model does
not have any non-parametric terms, so that we can estimate the coefficients
using maximum likelihood (standard errors are estimated from the usual
normal approximation at the mode).

Again we use the first 100 images in the set to estimate the parameters,
and keep the next 100 for model comparison. The fitted coefficient
for distance to the initial location is $\hat{\gamma}=-3.2$ (std.
err 0.1), and for distance to the center we find $\hat{\nu}=-1.6$
(std. err. 0.1). The value of $\gamma$ indicates a clear dependency
on initial location: everything else being equal, at a distance of
half the width of the screen the intensity has dropped by a factor
5. To see if the presence of this dependency induces a bias in the
estimation of coefficients for the saliency map, we also fitted the
model without the initial location term (setting $\gamma$ to 0).

\begin{figure}
\begin{centering}
\includegraphics[scale=0.6]{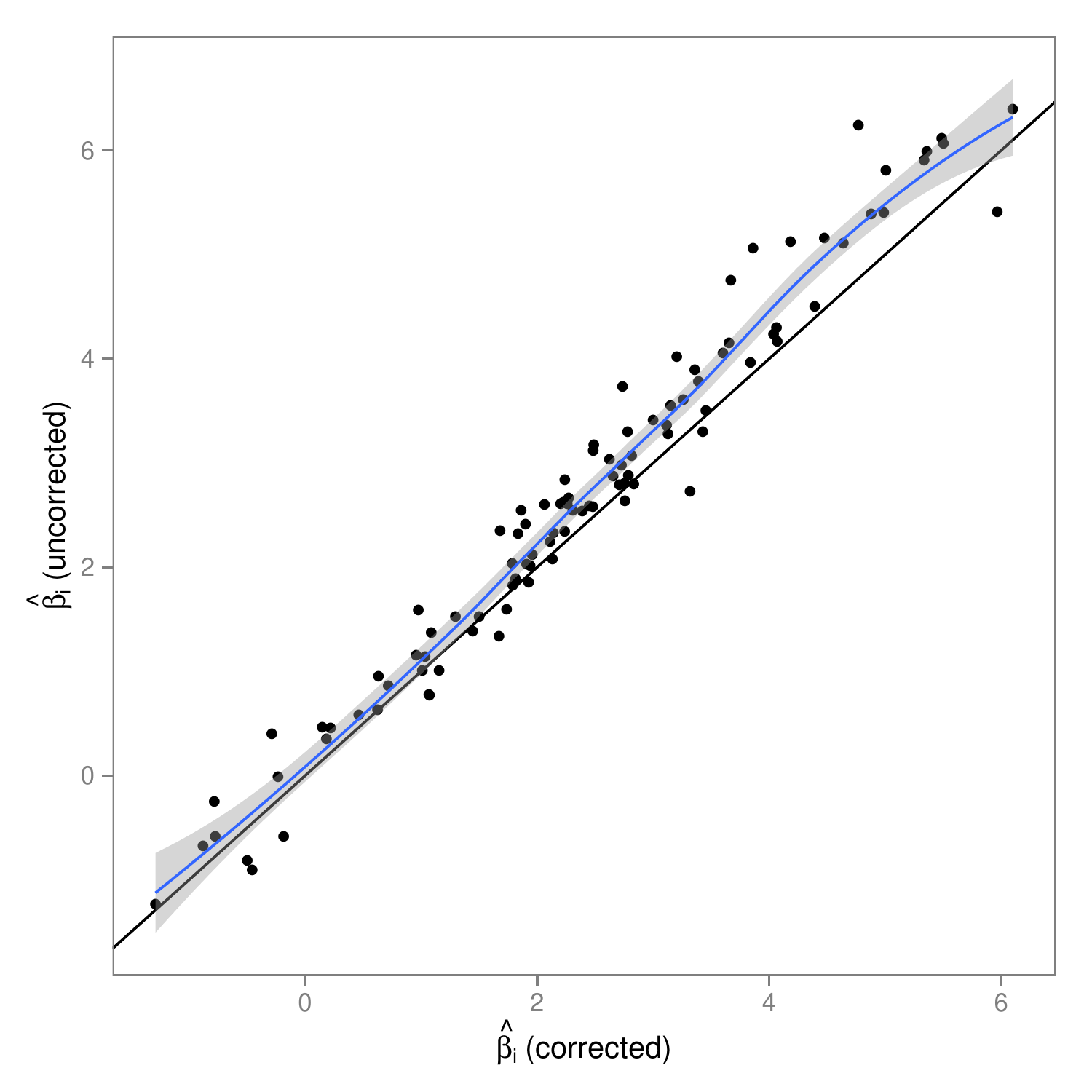} 
\par\end{centering}

\caption{Effect of correcting for dependence on the initial fixation location
(see Fig. \ref{fig:dependence-on-initial-fixation}). We compare the
estimated value of $\beta_{i}$ using the model of Equation \ref{eq:initial-fixation-model},
to coefficients estimated from a constrained model that does not include
the initial fixation as a covariate. The dots are coefficient pairs
corresponding to an image, and in blue we have plotted a trend line.
The diagonal line corresponds to equality. Although we do find evidence
of dependence on initial fixation location (see text), it does not
seem to cause any estimation bias: if anything the coefficients associated
with saliency are slightly higher when estimated by the uncorrected
model.\label{fig:Effect-of-correcting-coefficients}}
\end{figure}

We compared the estimated values for $\beta_{i}$ in both models,
and show the results on Figure \ref{fig:Effect-of-correcting-coefficients}:
the differences are minimal, although as we have argued there is certainly
some amount of dependence on the initial fixation. The lack of an
observed effect on the estimation of $\beta$ is probably due to the
fact that different observers fixate initially in different locations,
and that the dependency effectively washes out in the aggregate. An
interesting observation is that in the reduced model the coefficient
associated with distance to the center is estimated at -4.1 (std.
err. 0.1), which is much larger than when distance to initial fixation
is included as a covariate. Since initial fixations are usually close
to the center, being close to the center is correlated with being
close to the initial fixation location, and part of the centrality
bias observed in the dataset might actually be better recast as dependence
on the initial location.

In this we have managed to capture a source of dependence between
fixations, while seemingly still saving the IPP assumption. We have
done so by positing conditional independence: in our improved model
all fixations in a sequence are independent given the initial one.
An alternative is to drop the independence assumption altogether and
use dependent point process models, in which the location of each
point depends on the location of its neighbours. These models are
beyond the scope of this paper, but they are discussed extensively
in \citet{Diggle:StatisticalAnalysisSpatialPointPatterns} and \citet{Illian:StatAnalysisSpatPointPatterns},
along with a variety of non-parametric methods that can diagnose interactions.

\section{Discussion}

We introduced spatial point processes, arguing that they provide a
fruitful statistical framework for the analysis of fixation locations
and patterns for researchers interested in eye movements. 
In our exposition we analyzed a low-level saliency model by \citet{IttiKoch:ComputationalModellingVisualAttention},
and we were able to show that---although the model had predictive
value on average---it had varying usefulness from one image to another.
We believe that the consequences of this problem for prediction are
under-appreciated: as we stated in Section \ref{sub:Including-a-spatial-bias},
when trying to predict fixations over an arbitrary image, this variability
in quality of the predictor leads to predictions that are necessarily
vague. Although insights like this one could be arrived at starting
from other viewpoints, they arise very naturally from the spatial
point process framework presented here. Indeed, older methods of analysis
can be seen as approximations to point process model, as we shall
see below.

Owing to the tutorial nature of the material, there are some important
issues we have so far set swept under the proverbial rug. The first
is the choice of log-additive decompositions: are there other options,
and should one use them? The second issue is that of causality \citep{Henderson:HumGazeControlRealWorldScenePerception}:
can our methods say anything about what drives eye movements? Finally,
we also need to point out that the scope of point process theory is
more extensive than what we have been able to explore in this article,
and the last part of the discussion will be devoted to other aspects
of the theory that could be of interest to eye movement researchers.

\subsection{Point processes versus other methods of analysis}

In Section \ref{sub:FixLocationLocalProp}, we described various methods
that have been used in the analysis of fixation locations. Many treat
the problem of determining links between image covariates and fixation
locations as a patch classification problem: one tries to tell from
the contents of an image patch whether it was fixated or not. In the
appendix (Section \ref{sub:The-patch-classification-problem}), we
show that patch classification has strong ties to point process models,
and under some specific forms can be seen as an approximation to point
process modelling. In a nutshell, if one uses logistic regression
to discriminate between fixated and non-fixated patches, then one
is effectively modelling an intensity function. This fact is somewhat
obscured by the way the problem is usually framed, but comes through
in a formal analysis a bit too long to be detailed here. This result
has strong implications for the logic of patch classification methods,
especially regarding the selection of control patches, and we encourage
interested readers to take a look at Section \ref{sub:The-patch-classification-problem}.
Point process theory also allows for a rigorous examination of earlier
methods. We take a look in the appendix at AROC values and area counts,
two metrics that have been used often in assessing models of fixations.
We ask for instance what the ideal model would be, according to the
AROC metric, if fixations come from a point process with a certain
intensity function. The answer turns out to depend on how the control
patches are generated, which is rather crucial to the correct interpretation
of the AROC metric. This result and related ones are proved in Section
\ref{sub:ROC-and-Area-counts}.

We expect that more work will uncover more links between existing
methods and the point process framework. One of the benefits of thinking
in terms of the more general framework is that many results have already
been proven, and many problems have already been solved. We strongly
believe that the eye movement community will be able to benefit from
the efforts of others who work on spatial data.

\subsection{Decomposing the intensity function}

Throughout the article we have assumed a log-additive form for our
models, writing the intensity function as

\begin{equation}
\lambda\left(x,y\right)=\exp\left(\sum\alpha_{i}v_{i}(x,y)\right)\label{eq:log-add-decomp}
\end{equation}

for a set of covariates $v_{1},\ldots,v_{n}$. This choice may seem
arbitrary - for example, one could use

\begin{equation}
\lambda\left(x,y\right)=\sum\alpha_{i}v_{i}(x,y)\label{eq:add-decomp}
\end{equation}

a type of mixture model similar to those used in \citet{Vincent:DoWeLookAtLights}.
Since $\lambda$ needs to be always positive, we would have to assume
restrictions on the coefficients, but in principle this decomposition
is just as valid. Both (\ref{eq:log-add-decomp}) and (\ref{eq:add-decomp})
are actually special cases of the following:

\[
\lambda\left(x,y\right)=\Phi\left(\sum\alpha_{i}v_{i}(x,y)\right)
\]

for some function $\Phi$ (analoguous to the inverse link function
in Generalised Linear Models, see \citealp{McCullaghNelderGLMs}).
In the case of Equation \ref{eq:log-add-decomp} we have $\Phi\left(x\right)=\exp\left(x\right)$
and in the case of \ref{eq:add-decomp} we have $\Phi\left(x\right)=x$.
Other options are available, for instance \citet{Park:ActiveLearningNeuralResponseFunc}
use the following function in the context of spike train analysis:

\[
\Phi\left(x\right)=\log(\exp\left(x\right)+1)
\]

which approximates the exponential for small values of $x$ and the
identity for large ones. Single-index models treat $\Phi$ as an unknown
and attempt to estimate it from the data \citep{McCullaghNelderGLMs}.
From a practical point of view the log-additive form we use is the
most convenient, since it makes for a log-likelihood function that
is easy to compute and optimise, and does not require restrictions
on the space of parameters. From a theoretical perspective, the log-additive
model is compatible with a view that sees the brain as combining multiple
interest maps $v_{1},v_{2},...$ into a master map that forms the
basis of eye movement guidance. The mixture model implies on the contrary
that each saccade comes from a roll of dice in which one chooses the
next fixation according to one of the $v_{i}$'s. Concretely speaking,
if the different interest maps are given by, e.g. contrast and edges,
then each saccade is either contrast-driven with a certain probability
or on the contrary edges-driven.

We do not know which model is the more realistic, but the question
could be addressed in the future by fitting the models and comparing
predictions. It could well be that the situation is actually even
more complex, and that the data are best described by both linear
and log-linear mixtures: this would be the case, for example, if occasional
re-centering saccades are interspersed with saccades driven by an
interest map.

\subsection{Causality}

We need to stress that the kind of modelling we have done here does
not address causality. The fact that fixation locations can be predicted
from a certain spatial covariate does not imply that the spatial covariate
causes points to appear. To take a concrete example, one can probably
predict the world-wide concentration of polar bears from the quantities
of ice-cream sold, but that does not imply that low demand for ice-cream
causes polar bears to appear. The same caveat apply in spatial point
process models as in regression modelling, see \citet{GelmanHill:DataAnalysisUsingRegression}.
Regression modelling has a causal interpretation only under very restrictive
assumptions.

In the case of determining the causes of fixations in natural images,
the issue may actually be a bit muddled, as different things could
equally count as causing fixations. Let us go back to polar bears
and assume that, while rather indifferent to ice cream, they are quite
partial to seals. Thus, the presence of seals is likely to cause the
appearance of polar bears. However, due to limitations inherent to
the visual system of polar bears, they cannot tell between actual
seals and giant brown slugs. The presence of giant brown slugs then
also causes polar bears to appear. Both seals and giant brown slugs
are valid causes of the presence of polar bears, in the counterfactual
sense: no seal, no polar bear, no slug, no polar bear either. A more
generic description at the algorithmic level is that polar bears are
drawn to anything that is brown and has the right aspect ratio. At
a functional level, polar bears are drawn to seals because that is
what the polar bear visual system is designed to do.

The same goes for saccadic targeting: an argument is sometimes made
that fixated and non-fixated patches only differ in some of their
low-level statistics because people target objects, and the presence
of objects tend to cause these statistics to change \citep{Nuthmann:ObjectBasedAttentionalSelection}.
While the idea that people want to look at objects is a good \emph{functional
}account, at the algorithmic level they may try to do so by targeting
certain local statistics. There is no confounding in the usual sense,
since both accounts are equally valid but address different questions:
the first is algorithmic (how does the visual system choose saccade
targets based on an image?) and the other one teleological (what is
saccade targeting supposed to achieve?). Answering either of these
questions is more of an experimental problem than one of data analysis,
and we cannot---and do not want to---claim that point process modelling
is able to provide anything new in this regard.

\subsection{Scope and limitations of point processes}

Naturally, there is much we have left out, but at least we would like
to raise some of the remaining issues. First, we have left the temporal
dimension completely out of the picture. Nicely, adding a temporal
dimension in point process models presents no conceptual difficulty;
and we could extend the analyses presented here to see in detail whether,
for example, low-level saliency predicts earlier fixations better
than later ones. We refer the reader to \citet{RodriguesDiggle:BayesEstPredSpatTempLGCoxProc}
and \citet{ZammitMangion:PointProcModellingAfghanWarDiary} for recent
work in this direction.

Second, in this work we have considered that a fixation is nothing
more than a dot: it has spatial coordinates and nothing more. Of course,
this is not true: a fixation lasted a certain time, during which particular
fixational eye movements occured, etc. Studying fixation duration
is an interesting topic in its own right, because how long one fixates
might be tied to the cognitive processes at work in a task \citep{Nuthmann:CRISP}.
There are strong indications that when reading, gaze lingers longer
on parts of text that are harder to process. Among other things, the
less frequent a word is, the longer subjects tend to fixate it \citep{Kliegl:TrackingTheMindDuringReading}.
Saliency is naturally not a direct analogue of word frequency, but
one might nonetheless wonder whether interesting locations are also
fixated longer. We could take our data to be fixations coupled with
their duration, and we would have what is known in the spatial statistics
literature as a \emph{marked point process}. Marked point processes
could be of extreme importance to the analysis of eye movements, and
we refer the reader to \citet{Illian:HierarchicalSPPAnalysisPlantCommunity}
for some ideas on this issue.

Third, another limitation we need to state is that the point process
models we have described here do not deal very well with high measurement
noise. We have assumed that what is measured is an actual fixation
location, and not a noisy measurement of an actual measurement location.
In addition, the presence of noise in the oculomotor system means
that actual fixation location may not be the intended one, which of
course adds an intractable source of noise to the measurements. Issues
do arise when the scale of measurement error is larger than the typical
scale at which spatial covariates change. Although there are theoretical
solutions to this problem (involving mixture models), they are rather
cumbersome from a computational point of view. An less elegant work-around
is to blur the covariates at the scale of measurement error.

Finally, representing the data as a set of locations may not always
be the most appropriate way to think of the problem. In a visual search
task for example, a potentially useful viewpoint would be to think
of a sequence of fixations as covering a certain area of the stimulus.
This calls for statistical models that address random shapes rather
than just random point sets, an area known as stochastic geometry
\citep{Stoyan:StochasticGeometryAndItsApplications}, and in which
point processes play a central role, too.

\section{Appendices}

\subsection{The messy details of spatial statistics, and how to get around them }

The field of applied spatial statistics has evolved into a powerful
toolbox for the analysis of eye movements. There are, however, two
main hurdles in terms of accessibility. First, compared to eye-movement
research, the more traditional application fields (ecology, forestry,
epidemiology) have a rather separate set of problems. Consequently,
textbooks (e.g., \citet{Illian:StatAnalysisSpatPointPatterns}), focus
on non-Poisson processes, since corresponding problems often involve
mutual interactions of points, e.g., how far trees are from one another
and whether bisons are more likely to be in groups of three than all
by themselves. Such questions have to do with the second-order properties
of point processes, which express how points attract or repel one
another. The formulation of point process models with non-trivial
second-order properties, however, requires rather sophisticated mathematics,
so that the application to eye-movement data is no longer straight-forward.

Second, while the formal properties of point process models are well-known,
practical use is hindered by computational difficulties. A very large
part of the literature focuses on computational techniques (maximum
likelihood or Bayesian) for fitting point process models. Much progress
has been made recently (see, among others, \citealp{HaranTierney:AutomatingMCMCSpatialModels},
or \citealp{Rue:INLA}). Since we these technical difficulties might
not be of direct interest to most eye-movement researchers, we developed
a toolkit for the R environment that attempts to mathematical details
under the carpet. We build on one of the best techniques available
(INLA) \citep{Rue:INLA} to provide a generic way to fit multiple
point process models without worrying too much about the underlying
mathematics. The toolkit and a manual in the form of the technical
report has been made available for download on the first author's
webpage.

\subsection{Gaussian processes and Gauss-Markov processes\label{sub:Intro-to-GPs}}

Gaussian Processes (GPs) and related methods are tremendously useful
but not the easiest to explain. We will stay here at a conceptual
level, computational details can be found in the monograph of \citet{RasmussenWilliamsGP}.

\begin{figure}
\includegraphics[scale=0.5]{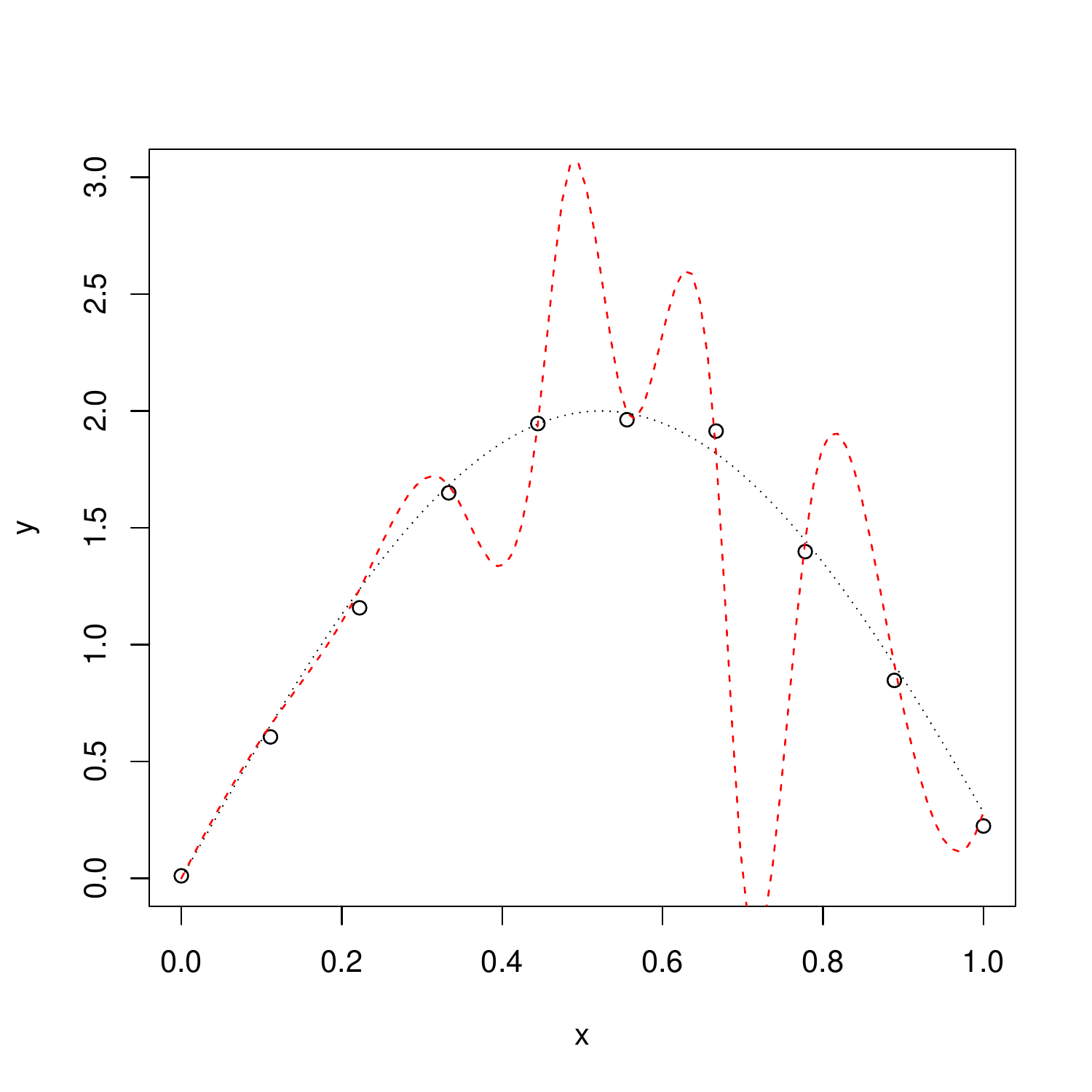}

\caption{A non-parametric regression problem. We have measured some output
$y$ for 10 different values of some covariate $x$. We plot these
data as open circles. Our assumption is that $y=f(x)+\epsilon$, where
$\epsilon$ is zero-mean noise. We wish to infer the underlying function
$f$ from the data without assuming a parametric form for $f$. The
two functions shown as dotted curves both have the same likelihood
- they are equally close to the data. In most cases the function in
red will be a far worse guess than the one in black. We need to inject
that knowledge into our inference, and this can be done by imposing
a prior on possible latent functions $f$. This can be done using
a GP. \label{fig:non-parametric-regression}}
\end{figure}

Rather than directly state how we use GPs in our models, we start
with a detour on non-parametric regression (see Figure \ref{fig:non-parametric-regression}),
which is were Gaussian processes are most natural. In non-parametric
regression, given the (noisy) values of a function $f\left(x\right)$
measured at points $x_{1},\ldots,x_{n}$, we try to infer what the
values of $f$ are at other points. \emph{Interpolation }and \emph{extrapolation
}can be seen as special cases of non-parametric regression - ones
where noise is negligible. The problem is non-parametric because we
do not wish to assume that $f(x)$ has a known parametric form (for
example, that $f$ is linear).

For a statistical solution to the problem, we need a likelihood, and
usually it is assumed that $y_{i}|x_{i}\sim\N\left(f\left(x_{i}\right),\sigma^{2}\right)$
which corresponds to observing the true value corrupted by Gaussian
noise of variance $\sigma^{2}$. This is not enough, since there are
uncountably many functions $f$ that have the same likelihood, namely
all those that have the same value at the sampling points $x_{1},\ldots,x_{n}$
(Fig. \ref{fig:non-parametric-regression}).

\begin{figure}
\begin{centering}
\subfloat[GPs can be specified through their \emph{covariance function $k(x,x')$.
}The covariance function expresses the following: if we were to measure
$f$ at $x$ and $x'$, how similar would we expect $f(x)$ and $f(x')$
to be? The classical \emph{Gaussian }or \emph{Matern }families of
covariance functions impose that expected similarity go down with
the distance between $x$ and $x'$. \textbf{a. }Two Gaussian covariance
functions with different length-scales: correlation drops faster for
one than the other (shorter length-scale). \textbf{b. }Samples from
the corresponding GPs: we see that a shorter length-scale leads to
less smooth functions.\label{fig:GP-cov-function}]{\includegraphics[scale=0.5]{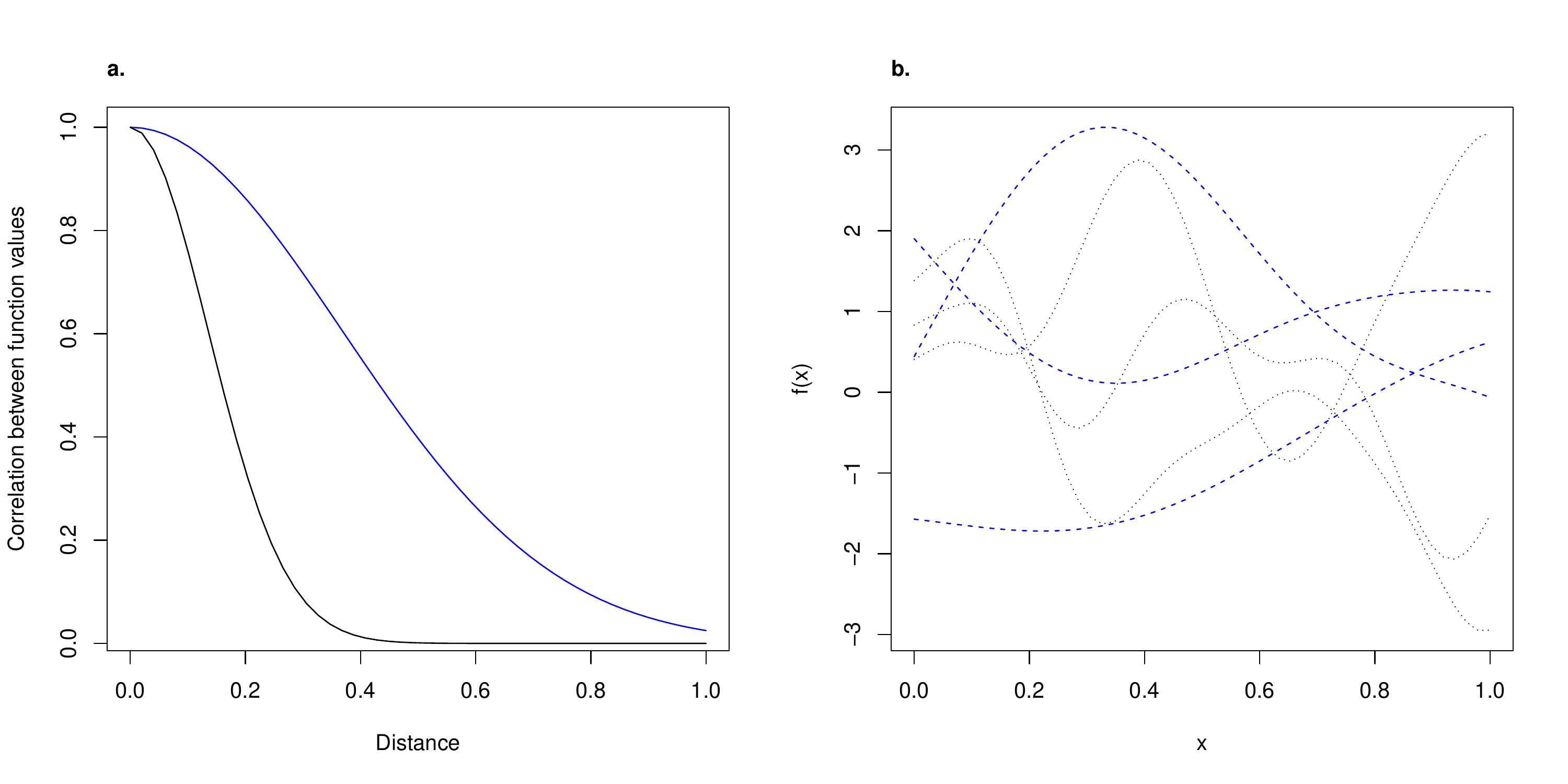}

}
\par\end{centering}

\begin{centering}
\subfloat[Bayesian update of a Gaussian Process. We start with a prior distribution
$p(f)$ over possible functions, then update the prior with data $\mathbf{y}$,
to get a posterior $p(f|\mathbf{y})\propto p(\mathbf{y}|f)p(f)$.
The posterior distribution is also a probability distribution, but
relative to the prior it is concentrated over the functions that are
likely given the data. On this Figure we show the data from Figure
\ref{fig:non-parametric-regression}, along with functions sampled
from the posterior distribution. ]{\includegraphics[scale=0.5]{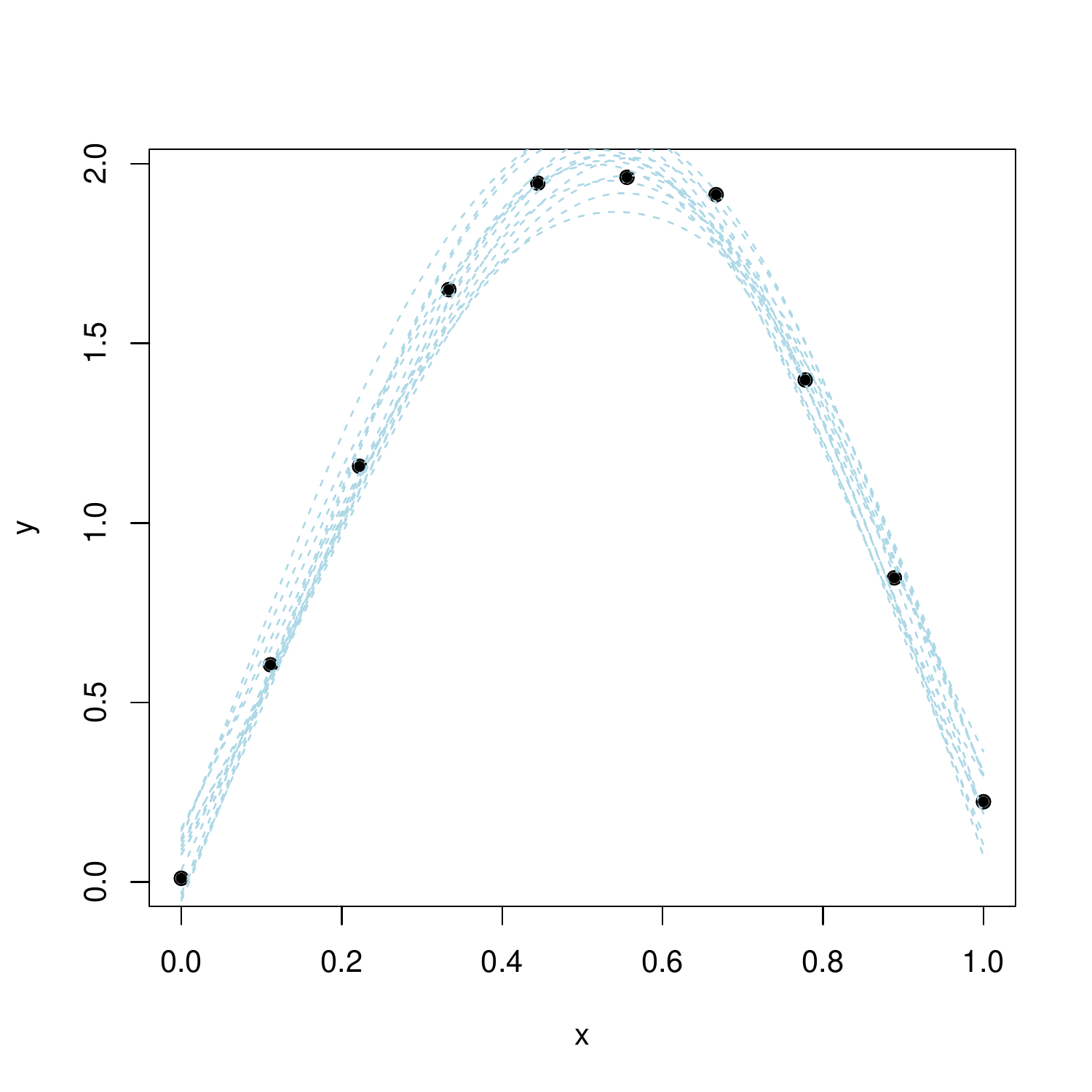}

}
\par\end{centering}

\caption{Gaussian Processes in the context of non-parametric regression. IPPs
are distributions over point sets, GPs are distributions over functions.
They can be used to specify a preference for ``reasonable'' functions.\label{fig:GPs-nonparam-reg}}
\end{figure}

Thus, we need to introduce some constraints. Parametric methods constrain
$f$ to be in a certain class, and can be thought of as imposing ``hard''
constraints. Nonparametric methods such as GP regression impose \emph{soft
}constraints, by introducing an a priori probability on possible functions
such that reasonable functions are favoured (Fig. \ref{fig:non-parametric-regression}
and \ref{fig:GPs-nonparam-reg}). In a Bayesian framework, this works
as follows. What we are interested in is the posterior probability
of $f$ given the data, which is as usual given by $p(f|\mathbf{y})\propto p(\mathbf{y}|f)p(f)$.
As we mentioned above $p(\mathbf{y}|f)=\prod_{i=1}^{n}\N\left(y_{i}|f(x_{i}),\sigma^{2}\right)$
is equal for all functions that have the same values at the sampled
points $x_{1},\ldots,x_{n}$, so what distinguishes them in the posterior
is how likely they are a priori---which is, of course, provided by
the prior distribution $p(f)$.

\begin{figure}
\begin{centering}
\subfloat[Stochastic processes can be used to generate random functions: here
we show three realisations from a Wiener process. The Wiener process
is a continuous analogue of the random walk. Although usually presented
as representing the movement of a particle, one can think of the path
taken by the Wiener process as a function $y(t)$, and therefore of
the Wiener process as generating a probability distribution over functions.
The Wiener process is a GP, but GPs used in practice generate much
smoother functions (see Figure \ref{fig:GPs-nonparam-reg} below).
\label{fig:Samples-Wiener-Process}]{\includegraphics[scale=0.35]{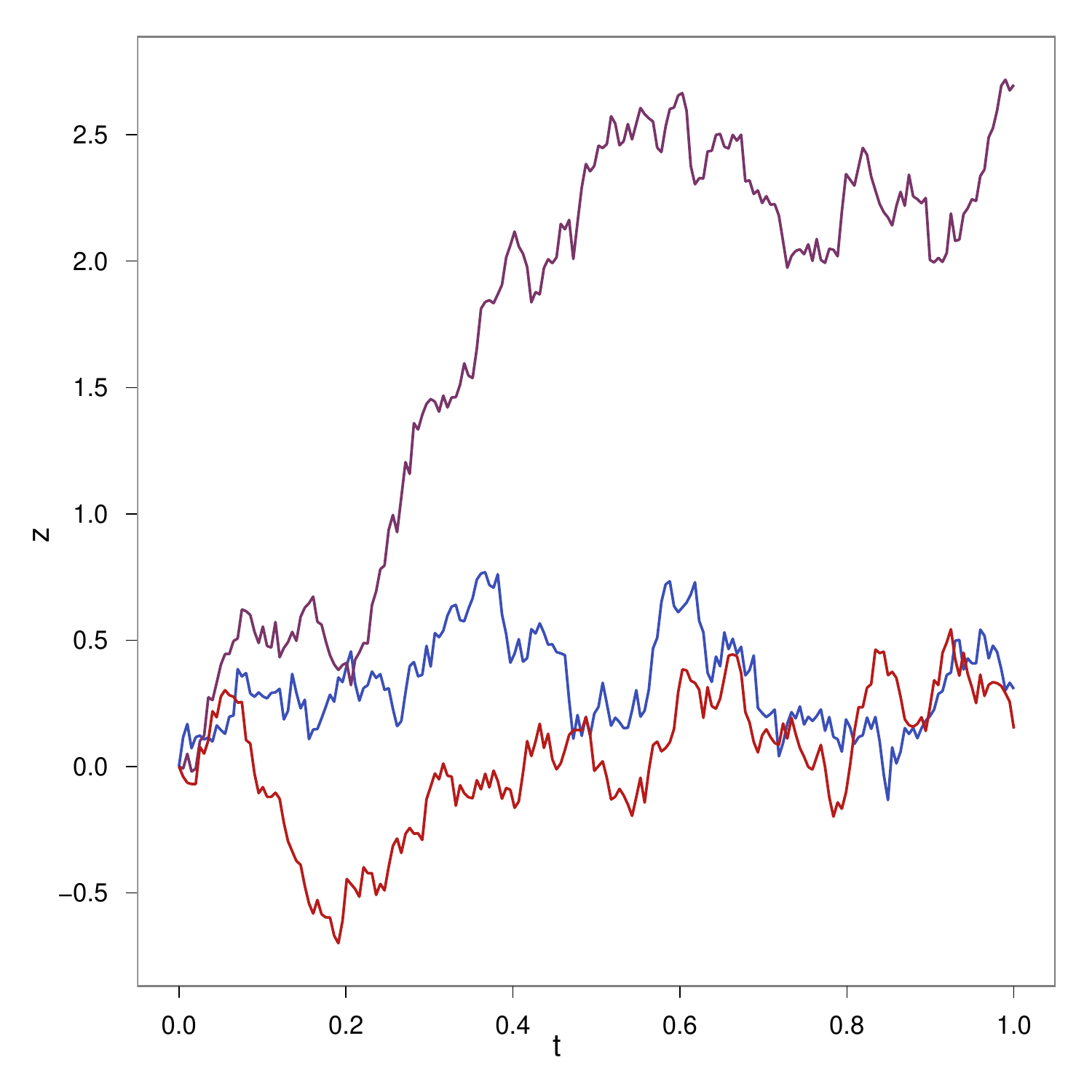}

}
\par\end{centering}

\begin{centering}
\subfloat[We generated 200 realisations of the Wiener process, and plot their
value at time $t=0.5$ against their value after either a small time
lag ($\delta=0.02$), or a larger time lag ($\delta=0.2$). The smaller
the time lag, the more these values are correlated. In general, this
property is reflected in the \emph{covariance function }of the GP.
\label{fig:Cov-Wiener-process}]{\includegraphics[scale=0.4]{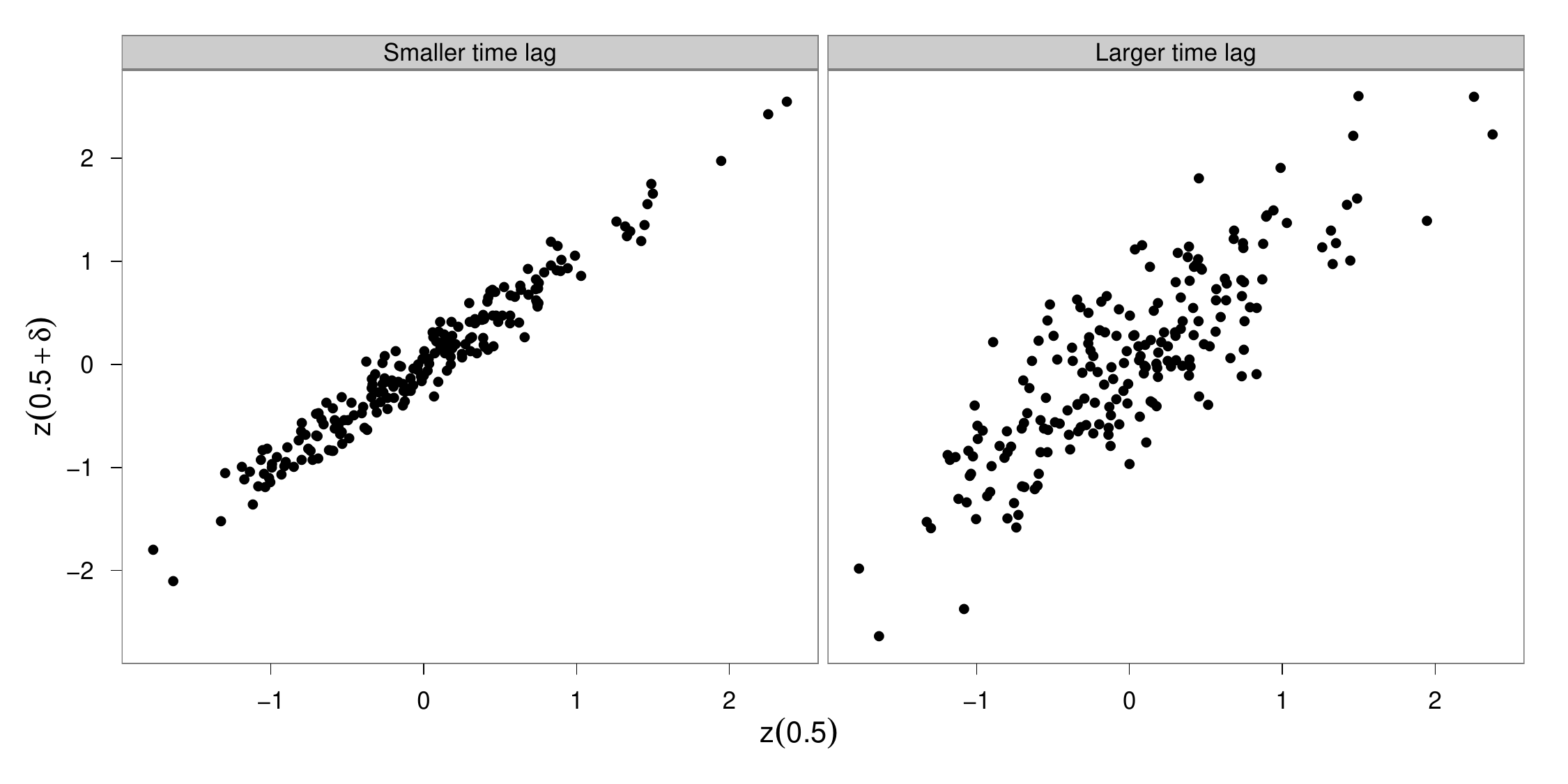}

}
\par\end{centering}

\caption{The Wiener Process, a member of the family of Gaussian Processes.\label{fig:Wiener-process}}
\end{figure}

How to formulate $p(f)$? We need a probability distribution that
is defined over a space of functions. The idea of a process that generates
random functions may not be as unfamiliar as it sounds: a Wiener process,
for example, can be interpreted as generating random functions (Fig.
\ref{fig:Samples-Wiener-Process}). A Wiener process is a diffusion:
it describes the random motion of a particle over time. To generate
the output of a Wiener process, you start at time $t_{0}$ with a
particle at position $z(t_{0})$, and for each infinitesimal time
increment you move the particle by a random offset, so that over time
you generate a ``sample path'' $z(t)$.

This sample path might as well be seen as a function, just like the
notation $z(t)$ indicates, so that each time one runs a Wiener process,
one obtains a different function. This distribution will probably
not have the required properties for most applications, since samples
from a Wiener process are much too noisy - they generate functions
that look very rough and jagged. The Wiener process is however a special
case of a GP, and this more general family has some much more nicely-behaved
members.

A useful viewpoint on the Wiener process is given by how successive
values depend on each other. Suppose we simulate many sample paths
of the Wiener Process, and each time measure the position at time
$t_{a}$ and $t_{b}$, so that we have a collection of $m$ samples
$\left\{ \left(z_{1}\left(t_{a}\right),z_{1}\left(t_{b}\right)\right),\ldots,\left(z_{m}\left(t_{a}\right),z_{m}\left(t_{b}\right)\right)\right\} $.
It is clear that $z\left(t_{a}\right)$ and $z\left(t_{b}\right)$
are not independent: if $t_{a}$ and $t_{b}$ are close, then $z(t_{A})$
and $z\left(t_{B}\right)$ will be close too. We can characterise
this dependence using the covariance between these two values: the
higher the covariance, the more likely $z\left(t_{a}\right)$ and
$z\left(t_{b}\right)$ are to be close in value. Figure \ref{fig:Cov-Wiener-process}
illustrates this idea.

If we could somehow specify a process such that the correlation between
two function values at different places does not decay too fast with
the distance between these two places, then presumably the process
would generate rather smooth functions. This is exactly what can be
achieved in the GP framework. The most important element of a GP is
the covariance function $k(x,x')$%
\footnote{A GP also needs a mean function, but here we will assume that the
mean is uniformly 0. See \citet{RasmussenWilliamsGP} for details.%
}, which describes how the covariance between two function values depend
on where the function is sampled: $k(x,x')=\mbox{Cov}\left(f(x),f(x')\right)$.

We now have the necessary elements to define a GP formally. A GP with
mean 0 and covariance function $k(x,x')$ is a distribution on the
space of functions of some input space $\mathcal{X}$ into $\mathbb{R}$,
such that for every set of $\left\{ x_{1},\ldots,x_{n}\right\} $,
the sampled values $f(x_{1}),\ldots,f(x_{n})$ are such that

\begin{eqnarray*}
f(x_{1}),\ldots,f(x_{n}) & \sim & \N\left(0,\mathbf{K}\right)\\
\mathbf{K}_{ij} & = & k(x_{i},x_{j})
\end{eqnarray*}

In words, the sampled values have a multivariate Gaussian distribution
with a covariance matrix given by the covariance function $k$. This
definition should be reminiscent of that of the IPP: here too we define
a probability distribution over infinite-dimensional objects by constraining
every finite-dimensional marginal to have the same form.

A shorthand notation is to write that

\[
f\sim\mathcal{GP}\left(0,k\right)
\]

and this is how we define our prior $p(f)$.

Covariance functions are often chosen to be Gaussian in shape%
\footnote{For computational reasons we favour here the (also very common) Matern
class of covariance functions, which leads to functions that are less
smooth than with a squared exponential covariance.%
} (sometimes called the ``squared exponential'' covariance function,
to avoid giving Gauss overly much credit):

\[
k(x,x')=\nu\exp\left(-\lambda\left(x-x'\right){}^{2}\right)
\]

It is important to be aware of the roles of the hyperparameters, here
$\nu$ and $\lambda$. Since $k(x,x)=\mbox{Var}\left(f(x)\right)$,
we see that $\nu$ controls the marginal variance of $f$. This gives
the prior a scale: for example, if $\nu=1$, the variance of $f(x)$
is 1 for all $x$, and because $f(x)$ is normally distributed this
implies that we do not expect $f$ to take values much larger than
3 in magnitude. $\lambda$ plays the important role of controlling
how fast we expect $f\left(x\right)$ to vary: the greater $\lambda$
is, the faster the covariance decays. What this implies is for very
low values of $\lambda$ we expect $f$ to be locally almost constant,
for very large values we expect it to vary much faster (Fig. \ref{fig:GP-cov-function}).
In practice it is often better (when possible) not to set the hyperparameters
to pre-specified values, but infer them also from the data (see \citealp{RasmussenWilliamsGP}
for details).

One of the concrete difficulties with working with Gaussian Processes
is related to the need to invert large covariance matrices when performing
inference. Inverting a large, dense matrix is an expensive operation,
and a lot of research has gone into finding ways of avoiding that
step. One of the most promising is to approximate the Gaussian Process
such that the \emph{inverse }covariance matrix (the precision matrix)
is sparse, which leads to large computational savings. Gauss-Markov
processes are a class of distributions with sparse inverse covariance
matrices, and the reader may consult \citet{RueHeld:GMRFTheoryApplications}
for an introduction.

\subsection{Details on Inhomogeneous Poisson Processes\label{sub:Details-on-IPPs}}

We give below some details on the likelihood function for inhomogeneous
Poisson processes (IPPs), as well as the techniques we used for performing
Bayesian inference.

\subsubsection{The likelihood function of an IPP}

An IPP is formally characterised as follows: given a spatial domain
$\Omega$, e.g here $\Omega=[0,1]^{2}$, and an intensity function
$\lambda:\,\Omega\rightarrow\mathbb{R}^{\text{+}}$ then an IPP is
a probability distribution over finite subsets $S$ of $\Omega$ such
that, for all sets $\mathcal{D}\in\Omega$, 
\begin{equation}
\left|S\cap\mathcal{D}\right|\sim Poi\left(\int_{\mathcal{D}}\lambda\left(x\right)\mbox{d}x\right)\label{eq:IPP-poisson-marginal-property}
\end{equation}

$\left|S\cap\mathcal{D}\right|$ is short-hand for the cardinal of
$S\cap\mathcal{D}$, the number of points sampled from the process
that fall in region $\mathcal{D}$. Note that in IPP, for disjoint
subsets $\mathcal{D}_{1},\ldots,\mathcal{D}_{r}$ the distributions
of $\left|S\cap\mathcal{D}_{1}\right|,\ldots,\left|S\cap\mathcal{D}_{r}\right|$
are independent%
\footnote{In other words, knowing how many fixations there were on the upper
half of the screen should not tell you anything about how many there
were in the lower half. This might be violated in practice but is
not a central assumption for our purposes. %
}.

For purposes of Bayesian inference, we need to be able to compute
the likelihood, which is the probability of the sampled point set
$S$ viewed as a function of the intensity function $\lambda\left(\centerdot\right)$.
We will try to motivate and summarize the necessary concepts without
a rigorous mathematical derivation, interested readers should consult
\citet{Illian:StatAnalysisSpatPointPatterns} for details.

We note first that the likelihood can be approximated by \emph{gridding
}the data: we divide $\Omega$ into a discrete set of regions $\Omega_{1},\ldots,\Omega_{r}$,
and count how many points in $S$ fell in each of these regions. The
likelihood function for the gridded data is given directly by Equation
\ref{eq:IPP-poisson-marginal-property} along with the independence
assumption: noting $k_{1},\ldots,k_{r}$ the bin counts we have

\begin{eqnarray}
p\left(k_{1},\ldots,k_{r}|\lambda\right) & = & \prod_{j=1\ldots r}\frac{\left(\lambda_{j}\right)^{k_{j}}}{k_{j}!}\exp\left(-\lambda_{j}\right)\label{eq:Poisson-counts}\\
\lambda_{j} & = & \int_{\Omega_{j}}\lambda\left(x\right)\mbox{d}x\nonumber 
\end{eqnarray}

Also, since $\Omega_{1},\ldots,\Omega_{r}$ is a partition of $\Omega$,
$\prod\exp\left(-\lambda_{j}\right)=\exp(-\sum\lambda_{j})=\exp\left(-\int_{\Omega}\lambda\left(x\right)\mbox{d}x\right)$.

As we make the grid finer and finer we should recover the true likelihood,
because ultimately if the grid is fine enough for all instance and
purposes we will have the true locations. As we increase the number
of grid points $r$, and the area of each region $\Omega_{j}$ shrinks,
two things will happen: 
\begin{itemize}
\item the counts will be either 0 (in the vast majority of empty regions),
or 1 (around the locations $s_{1},\ldots,s_{n}$ of the points in
$S$). 
\item the integrals $\int_{\Omega_{j}}\lambda\left(x\right)\mbox{d}x$ will
tend to $\lambda\left(x_{j}\right)\mbox{d}x$, with $x_{j}$ any point
in region $\Omega_{j}$. In dimension 1, this corresponds to saying
that the area under a curve can be approximated by height x length
for small intervals. 
\end{itemize}
Injecting this into Equation (\ref{eq:Poisson-counts}), in the limit
we have:

\begin{equation}
p(S|\lambda\left(\cdot\right))=\frac{1}{n}\left\{ \prod_{i=1}^{n}\lambda\left(s_{i}\right)\mbox{d}x\right\} \exp\left(-\int_{\Omega}\lambda\left(x\right)\mbox{d}x\right)\label{eq:likelihood-function-IPP}
\end{equation}

Since the factors $dx$ and $n^{-1}$ are independent of $\lambda$
we can neglect them in the likelihood function.

\subsubsection{Conditioning on the number of datapoints, computing predictive distributions\label{sub:Conditioning-on-n}}

The Poisson process has a remarkable property \citep{Illian:StatAnalysisSpatPointPatterns}:
conditional on sampling $n$ points from a Poisson process with intensity
function $\lambda\left(x,y\right)$, these $n$ points are distributed
independently with density 
\[
\bar{\lambda}\left(x,y\right)=\frac{\lambda\left(x,y\right)}{\int\lambda\left(x,y\right)\mbox{d}x\mbox{d}y}
\]

Intuitively speaking, this means that if you know the point process
produced 1 point, then this point is more likely to be where intensity
is high.

This property is the direct analogue of its better known discrete
variant: if $z_{1},z_{2},\ldots,z_{n}$ are independently Poisson
distributed with mean $\lambda_{1},\ldots,\lambda_{n}$, then their
joint distribution conditional on their sum $\sum z_{i}$ is multinomial
with probabilities $\pi_{i}=\frac{\lambda_{i}}{\sum\lambda_{j}}$.
Indeed, the continuous case can be seen as the limit case of the discrete
case.

We bring up this point because it has an important consequence for
prediction. If the task is to predict where the 100 next points are
going to be, then the relevant predictive distribution is:

\begin{equation}
p(S|S\mbox{ has size }n)=\prod_{i=1}^{n}\frac{\lambda\left(x_{i},y_{i}\right)}{\int\lambda\left(x,y\right)\mbox{d}x\mbox{d}y}\label{eq:distribution-n-known}
\end{equation}

where $S$ is a point set of size $n$, whose points have $x$ coordinates
$x_{1},\ldots,x_{n}$ and $y$ coordinates $y_{1},\ldots,y_{n}$ .
Equation \ref{eq:distribution-n-known} is the right density to use
when evaluating the predictive abilities of the model with $n$ known
(for example if one wants to compute the predictive deviance).

In the main text we had models of the form:

\[
\log\lambda_{i}\left(x,y\right)=\eta_{i}\left(x,y\right)=\alpha_{i}+\beta_{i}m_{i}(x,y)
\]

and we saw that when predicting data for a new image $j$ we do not
know the values of $\alpha_{j}$ and $\beta_{j}$, and need to average
over them. The good news is that when $n$ is known we need not worry
about the intercept $\alpha_{j}$: all values of $\alpha_{j}$ lead
to the same predictive distribution, because $\alpha_{j}$ disappears
in the normalisation in Equation \ref{eq:distribution-n-known}. Given
a distribution $p(\beta)$ for possible slopes, the predictive distribution
is given by:

\begin{eqnarray*}
p(S_{j}|S_{j}\mbox{ has size }n) & = & \int p(\beta_{j})\prod_{i=1}^{n}\frac{\exp\left(\beta_{j}m_{j}(x_{i},y_{i})\right)}{\int\exp\left(\beta_{j}m_{j}(x,y)\right)\mbox{d}x\mbox{d}y}\mbox{d}\beta_{j}
\end{eqnarray*}

It is important to realise that the distribution above does \emph{not
}factorise over points, unlike (\ref{eq:distribution-n-known}) above.
Computation requires numerical or Monte Carlo integration over $\beta$
(as far as we know).

\subsection{Approximations to the likelihood function}

One difficulty immediately arises when considering Equation (\ref{eq:likelihood-function-IPP}):
we require the integral$\int_{\Omega}\lambda\left(x\right)\mbox{d}x$.
While not a problem when $\lambda\left(\cdot\right)$ has some convenient
closed form, in the cases we are interested in $\lambda(x)=\exp\left(\eta\left(x\right)\right)$,
with $\eta\left(\cdot\right)$ a GP sample. The integral is therefore
not analytically tractable. A more fundamental difficulty is that
the posterior distribution $p(\lambda\left(\cdot\right)|S)$ is over
an infinite-dimensional space of functions - how are we to represent
it?

All solutions use some form of discretisation. A classical solution
is to use the approximate likelihood obtained by binning the data
(Eq. \ref{eq:Poisson-counts}), which is an ordinary Poisson count
likelihood. The bin intensities $\lambda_{j}=\int_{\Omega_{j}}\lambda\left(x\right)\mbox{d}x$
are approximated by assuming that bin area is small relative to the
variation in $\lambda\left(\cdot\right)$, so that: 
\[
\lambda_{j}=\lambda\left(x_{j}\right)\left|\Omega_{j}\right|
\]

with $|\Omega_{j}|$ the area of bin $\Omega_{j}$ and \textbf{$x_{j}$}.
The approximate likelihood then only depends on the value of $\lambda\left(\cdot\right)$
at bin centres, so that we can now represent the posterior as the
finite-dimensional distribution $p(\lambda_{1}\ldots\lambda_{r}|S)$.
In practice we target rather $p(\eta_{1}\ldots\eta_{r}|S)$, for which
the prior distribution is given by (see

\[
\eta(x_{1}),\ldots,\eta(x_{r})\sim\N\left(0,\mathbf{K}_{\theta}\right)
\]

Here $\mathbf{K}_{\theta}$ is the covariance matrix corresponding
to the covariance function $k_{\theta}\left(\cdot,\cdot\right)$,
and $\theta$ represents hyperparameters (e.g., marginal variance
and length-scale of the process).

A disadvantage of the binning approach is that fine gridding in 2D
requires many, many bins, which means that good spatial resolution
requires dealing with very large covariance (or precision) matrices,
slowing down inference.

Another solution, due to \citet{BermanTurner:ApproxPointProcessLikelihoods},
uses again the values of $\eta\left(\cdot\right)$ sampled at $r$
grid points, but approximates directly the original likelihood (\ref{eq:likelihood-function-IPP}).
The troublesome integral $\int_{\Omega}\lambda\left(x\right)\mbox{d}x$
is dealt with using simple numerical quadrature:

\[
\int_{\Omega}\lambda\left(x\right)\mbox{d}x\approx\sum w_{j}\exp\left(\eta\left(x_{j}\right)\right)
\]

where the $w_{j}$'s are quadrature weights. The values $\lambda\left(s_{i}\right)$
at the sampled points are interpolated from the known values at the
grid points:

\[
\lambda\left(s_{i}\right)=\exp\left(\sum_{j=1\ldots r}a_{ij}\eta\left(x_{j}\right)\right)
\]

the $a_{ij}$ are interpolation weights. Injecting into (\ref{eq:likelihood-function-IPP})
we have the approximate log-likelihood function:

\begin{equation}
\mathcal{L}\left(\eta\right)=\sum_{i,j}a_{ij}\eta\left(x_{j}\right)-\sum w_{j}\exp\left(\eta\left(x_{j}\right)\right)\label{eq:approx-IPP-likelihood-interp}
\end{equation}

This log-likelihood function is compatible with the INLA framework
for inference in latent Gaussian models (see \citealp{Rue:INLA} and
the website www.r-inla.org). The same log-likelihood function can
be also be used in the classical maximum-likelihood framework. Approximate
confidence intervals and p-values for coefficients can be obtained
from nested model comparisons, or using resampling techniques (these
techniques are explained in most statistics textbooks, including \citealp{Wasserman:AllOfStats}).

\subsection{Relating patch statistics to point process theory\label{sub:Relating-patch-statistics-to-PP-theory}}

As noted above much of the previous work in the literature has focused
on the statistics of image patches as a way of characterising the
role of local image features in saliency. In this section we will
show that these analyses can be related to point process modelling,
leading to new insights. First, patch classification task approximates
IPP models under certain assumptions. Second, the common practice
of measuring performance by the area under the ROC curve can be grounded
in point process assumptions. Third, the alternative procedure which
consists in measuring performance by the proportion of fixations in
the most salient region is in fact just a variant of the ROC procedure
when seen in the point process context.

\subsubsection{The patch classification problem as an approximate to point process
modeling\label{sub:The-patch-classification-problem}}

We show here that patch classification can be interpreted as an approximation
to point process modeling. Although we focus on the techniques used
in the eye movement literature, our analysis is very close in spirit
to that of \citet{Baddeley:SpatialLogisticRegAndChangeOfSupport}
on the spatial logistic regression methods used in Geographic Information
Systems. Note that \citet{Baddeley:SpatialLogisticRegAndChangeOfSupport}
contains in addition many interesting results not detailed here and
much more careful mathematical analysis.

Performing a classical patch statistics analysis involves collecting
for a set of $n$ fixations $n$ ``positive'' image patches around
the fixation locations and $n$ ``negative'' image patches around
$n$ control locations, and comparing the contents of the patches.
To avoid certain technical issues to do with varying intercepts (which
we discuss later), we suppose that all fixations come from the same
image. The usual practice is to compute some summary statistics on
each patch, for example its luminance and contrast, and we note $\mathbf{v}(x,y)\in\R^{d}$
the value of those summary statistics at location $x,y$. We will
see that $\mathbf{v}(x,y)$ plays the role of spatial covariates in
point process models. We assume that the $n$ control locations are
drawn uniformly from the image.

The next step in the classical analysis is to compare the conditional
distributions of $\mathbf{v}$ for positive and negative patches,
which we will denote $p(\mathbf{v}|s=1)$ and $p(\mathbf{v}|s=0)$.
If these conditional distributions are different, than they afford
us some way to tell between selected and non-selected locations based
on the contents of the patch. Equivalently, $p\left(s|\mathbf{v}\right)$,
the probability of the patch label given the covariates, is different
from the base rate of 50\% for certain values of $\mathbf{v}$. Patch
classification is concerned with modelling $p(s|\mathbf{v})$, a classical
machine learning problem that can be tackled via logistic regression
or a support vector machine, as in \citet{Kienzle:CenterSurroundPatternsOptimalPredictors}.

Under certain conditions, statistical analysis of the patch classification
problem can be related to point process modelling. Patch classification
can be related to \emph{spatial logistic regression}, which in turn
can be shown to be an approximation of the IPP model \citep{Baddeley:SpatialLogisticRegAndChangeOfSupport}.
We give here a simple proof sketch that deliberately ignores certain
technical difficulties associated with the smoothness (or lack thereof)
of $\mathbf{v}\left(x,y\right)$.

In spatial logistic regression, a set of fixations is turned into
binary data by dividing the domain into a large set of pixels $a_{1}\ldots a_{r}$
and defining a binary vector $z_{1}\ldots z_{r}$ such that $z_{i}=1$
if pixel $a_{i}$ contains a fixation and $z_{i}=0$ otherwise. The
second step is to regress these binary data onto the spatial covariates
using logistic regression, which implies the following statistical
model:

\begin{align}
p\left(\mathbf{z}|\bb\right) & =\prod_{i=1}^{r}\Lambda\left(\bb^{t}\mathbf{v}_{i}+\beta_{0}\right)^{z_{i}}\left(1-\Lambda\left(\bb^{t}\mathbf{v}_{i}+\beta_{0}\right)\right)^{1-z_{i}}\label{eq:spatial-logistic-model}\\
\Lambda\left(\eta\right) & =\frac{1}{1+e^{-\eta}}\label{eq:logistic-function}
\end{align}

Equation \ref{eq:logistic-function} is just the logistic function,
and the value $\mathbf{v}_{i}$ of the covariates at the $i-$th pixel
can be either the average value over the pixel or the value at the
center of the pixel. By the Poisson limit theorem, the independent
Bernoulli likelihood of Equation \ref{eq:spatial-logistic-model}
becomes that of a Poisson process as pixel size tends to 0. In this
limit the probability of observing any individual $z_{i}=1$ will
be quite small, so that $\eta_{i}=\bb^{t}\mathbf{v}_{i}+\beta_{0}$
will be well under 0 around the peak of the likelihood. For small
values of $\eta$, $\Lambda\left(\eta\right)\approx\exp\left(\eta\right)$,
so that:

\[
p\left(\mathbf{z}|\bb\right)\approx\prod_{i=1}^{r}\exp\left(\bb^{t}\mathbf{v}_{i}+\beta_{0}\right)^{z_{i}}\left(1-\exp\left(\bb^{t}\mathbf{v}_{i}+\beta_{0}\right)\right)^{1-z_{i}}
\]

We take the log of $p\left(\mathbf{z}|\bb\right)$ and split the indices
of the pixels according to the value of $z_{i}$: we note $\mathcal{I}^{+}$
the set of selected pixels (pixels such that $z_{i}=1$), $\mathcal{I}^{-}$
its complement. This yields:

\[
\log\, p\left(\mathbf{z}|\bb\right)\approx\sum_{\mathcal{I}^{+}}\left\{ \bb^{t}\mathbf{v}_{i}+\beta_{0}\right\} +\sum_{\mathcal{I}^{-}}\left\{ \log\left(1-\exp\left(\bb^{t}\mathbf{v}_{i}+\beta_{0}\right)\right)\right\} 
\]

We make use again of the fact that Bernoulli probabilities will be
small, which implies that $1-\exp\left(\bb^{t}\mathbf{v}_{i}+\beta_{0}\right)$
will be close to 1. Since $\log\left(x\right)\approx x-1$ for $x\approx1$,
the approximation can be further simplified to:

\[
\log\, p\left(\mathbf{z}|\bb\right)\approx\sum_{\mathcal{I}^{+}}\left\{ \bb^{t}\mathbf{v}_{i}+\beta_{0}\right\} -\sum_{\mathcal{I}^{-}}\exp\left(\bb^{t}\mathbf{v}_{i}+\beta_{0}\right)
\]

If the pixel grid is fine enough the second part of the sum will cover
almost every point, and will therefore be proportional to the integral
of $\bb^{t}\mathbf{v}\left(x,y\right)+\beta_{0}$ over the domain.
This shows that the approximate likelihood given in Equation (\ref{eq:spatial-logistic-model})
tends to the likelihood of an IPP (Eq. \ref{eq:likelihood-function-IPP})
with log intensity function $\log\lambda\left(x,y\right)=\bb^{t}\mathbf{v}\left(x,y\right)+\beta_{0}$,
which is exactly the kind used in this manuscript.

This establishes the small-pixel equivalence of spatial logistic regression
and IPP modeling. It remains to show that spatial logistic regression
and patch classification are in some cases equivalent.

In the case described above, one has data for one image only, collects
$n$ patches at random as examples of non-fixated locations, and then
performs logistic regression on the patch labels based on the patch
statistics $\mathbf{v}_{i}$. This is essentially the same practice
often used in spatial logistic regression, where people simply throw
out some of the (overabundant) negative examples at random. Throwing
out negative examples leads to a loss in efficiency, as shown in \citet{Baddeley:SpatialLogisticRegAndChangeOfSupport}
. It is interesting to note that under the assumption that fixations
are generated from a IPP with intensity $\lambda\left(x,y\right)=\bb^{t}\mathbf{v}\left(x,y\right)+\beta_{0}$,
giving us $n$ positive examples, and assuming that the $n$ negative
examples are generated uniformly, the logistic likelihood becomes
exact (this is a variant of lemma 12 in \citealp{Baddeley:SpatialLogisticRegAndChangeOfSupport}).
The odds-ratio that a point at location $x,y$ was one of the original
fixations is simply:

\begin{equation}
\log\,\frac{p(z=1|x,y)}{p(z=0|x,y)}=\log\,\frac{\lambda\left(x,y\right)}{A^{-1}}=\bb^{t}\mathbf{v}\left(x,y\right)+\beta_{0}+\log\left(A\right)\label{eq:logistic-regression-exact}
\end{equation}

Here $A$ is the total area of the domain. Equation \ref{eq:logistic-regression-exact}
shows in another way the close relationship between patch classification
and spatial point process models: patch classification using logistic
regression is a correct (but inefficient) model under IPP assumptions.

In actual practice patch classification involves a) combining patches
from multiple images and b) possibly different techniques for classification
than logistic regression. The first issue may be problematic, since
as we have shown above the coefficients relating covariates to intensity,
and certainly intercept values, may vary substantially from one image
to another. This can be fixed by changing the covariate matrix appropriately
in the logistic regression.

The second issue is that, rather than logistic regression, other classification
methods may be used : does the point process interpretation remain
valid? If one uses support vector machines, then the answer is yes,
at least in the limit of large datasets (SVM and logistic regression
are asymptotically equivalent, see \citealp{Steinwart:ConsistencySVMAndOtherRegKernelClass}):
the coefficients will differ only in magnitude. The same holds for
probit, complementary log-log or even least-squares regression: the
correct coefficients will be asymptotically recovered up to a scaling
factor. In actual practice, the difference between classification
methods is often small and all of them may equally be thought of as
approximating point process modeling.

\subsubsection{ROC performance measures, area counts, and minimum volume sets\label{sub:ROC-and-Area-counts}}

Many authors have used the area under the ROC curve as a performance
measure for patch classification. Another technique, used for example
in \citet{Torralba:ContextualGuidanceEyeMovements}, is to take the
image region with the 20\% most salient pixels and count the proportion
of fixations that occured in this region (a proportion over 20\% is
counted as above-chance performance). We show below that the two procedures
are related to each other and to point process models. The notion
of minimum volume regions will be needed repeatedly in the development
below, and we therefore begin with some background material.

\paragraph{Minimum-volume sets}

A minimum-volume set with confidence level $\alpha$ is the smallest
region that will contain a point with probability at least $\alpha$
(it extends the notion of a confidence interval to arbitrary domains,
see Fig. \ref{eq:maximum-probability-set}). Formally, given a probability
density $\pi(s)$ over some domain, the minimum volume set $F_{\alpha}$
is such that:

\begin{align}
F_{\alpha} & =\underset{F\in\mathcal{M}\left(\Omega\right)}{\mbox{argmin}}\mbox{Vol}\left(F\right)\label{eq:minimum-volume-set}\\
\mbox{subject to} & \int_{F}\pi\geq\alpha\nonumber 
\end{align}

where the minimisation is over all measurable subsets of the domain
$\Omega$ and $\mbox{Vol}\left(F\right)=\int_{F}1$ is the total volume
of set $F$. Intuitively speaking, the smaller the minimum-volume
sets of $\pi$, the less uncertainty there is: if 99\% of the probability
is concentrated in just 20\% of the domain, then $\pi$ is in some
sense 5 times more concentrated than the uniform distribution over
$\Omega$. In the case of Gaussian distributions the minimum volume
sets are ellipses, but for arbitrary densities they may have arbitrary
shapes. In \citet{NunezGarcia:LevelSetsMinimumVolumeSetsPDF} it is
shown that the family of minimum volume sets of $\pi$ is equal to
the family of contour sets of $\pi$: that is, every set $F_{\alpha}$
is equal to a set $\left\{ s\in\Omega|\pi\left(s\right)\geq\pi_{0}\right\} $
for some value $\pi_{0}$ that depends on $\alpha$%
\footnote{This is not true in the non-pathological cases where these sets are
not uniquely defined, for example in the case of the uniform distribution
where minimum-volume sets may be constructed arbitrarily. %
}. We can therefore measure the amount of uncertainty in a distribution
by looking at the volume of its contour sets, a notion which will
arise below when we look at optimal ROC performance.

\paragraph{ROC measures}

In ROC-based performance measures, the output of a saliency model
$m(x,y)$ is used to classify patches as fixated or not, and performance
classification is measured using the area of under the ROC curve.
The ROC curve is computed by varying a criterion $\xi$ and counting
the rate of False Alarms and Hits that result from classifying a patch
as fixated. In this section we relate this performance measure to
point process theory and show that: 
\begin{enumerate}
\item If non-fixated patches are sampled from a homogeneous PP, and fixated
patches from a IPP with intensity $\lambda\left(x,y\right)$, then
the optimal saliency model in the sense of the AUC metric is $\lambda(x,y)$
(or any monotonic transformation thereof). In this case the AUC metric
measures the precision of the IPP, i.e. how different from the uniform
intensity function $\lambda\left(x,y\right)$ is. 
\item If non-fixated patches are not sampled from a uniform distribution,
but from some other PP with intensity $\varphi\left(x,y\right)$,
then the optimal saliency model in the sense of the AUC metric is
no longer $\lambda\left(x,y\right)$ but $\frac{\lambda\left(x,y\right)}{\varphi\left(x,y\right)}$.
In other words, a saliency model could correctly predict fixation
locations but perform sub-optimally according to the AUC metric if
non-fixated locations are sampled from a non-uniform distribution
(for example when non-fixated locations are taken from other pictures). 
\item Since AUC performance is invariant to monotonic transformations of
$m(x,y)$, it says nothing about how intensity scales with $m(x,y)$. 
\end{enumerate}
In the following we will simplify notation by noting spatial locations
as $s=(x,y)$, and change our notation for functions accordingly ($m(s),\lambda(s)$,
etc.). We assume that fixated patches are drawn from a PP with intensity
$\lambda(s)$ and non-fixated patches from a PP with intensity $\varphi\left(s\right)$.
In ROC analysis locations are examined independently from one another,
so that all that matters are the normalised intensity functions (probability
densities for single points, see Section \ref{sub:Conditioning-on-n}).
Without loss of generality we assume that $\lambda$ and $\varphi$
integrate to 1 over the domain.

By analogy with psychophysics we define the task of deciding whether
a single, random patch $s$ was drawn from $\lambda$ (the fixated
distribution) or $\varphi$ (the non-fixated distribution) as the
Yes/No task. Correspondingly, the 2AFC task is the following: two
patches $s_{1},s_{2}$ are sampled random, one from $\lambda$ and
one from $\varphi$, and one must guess which of the two patches came
from $\lambda$. The Y/N task is performed by comparing the ``saliency''
of the patch $m(s)$ to a criterion $\xi$. The 2AFC task is performed
by comparing the relative saliency of $s_{1}$ and $s_{2}$: if $m(s_{1})>m(s_{2})$
the decision is that $s_{1}$ is the fixated location.

We will use the fact that the area under the ROC curve is equal to
2AFC performance \citep{GreenSwets:SDT}. 2AFC performance can be
computed as follows: we note $O=\left(1,0\right)$ the event corresponding
to $s_{1}\sim\lambda$ and $s_{2}\sim\varphi$. The probability of
a correct decision under the event $O=(1,0)$ is:

\begin{equation}
p_{c}=\int_{\Omega}\lambda(s_{1})\left\{ \int_{\Omega}\varphi\left(s_{2}\right)\I\left(m\left(s_{1}\right)>m(s_{2})\right)\mbox{d}s_{2}\right\} \mbox{d}s_{1}\label{eq:2AFC-prob-correct}
\end{equation}

where $\I\left(m\left(s_{1}\right)>m(s_{2})\right)$ is the indicator
function of the event that $s_{1}$ has higher saliency than $s_{2}$
(according to $m$). Note that the $O=\left(0,1\right)$ event is
exactly symmetrical, so that we do not need to consider both.

We first consider the case where non-fixated locations are drawn uniformly
($\varphi\left(s\right)=V^{-1}$, where $V$ is the area or volume
of the observation window), and ask what the optimal saliency map
$m$ is - in the sense of giving maximal 2AFC performance and therefore
maximal AUC. 2AFC can be viewed as a categorisation task over a space
of stimulus pairs, where the two categories are $O=(1,0)$ and $O=(0,1)$
in our notation. The so-called ``Bayes rule'' is the optimal rule
for categorisation \citep{DudaHartStork:PatternClassification}, and
in our case it takes the following form: answer $O=(1,0)$ if $p\left(O=(1,0)|s_{1},s_{2})\right)>p\left(O=(0,1)|s_{1},s_{2}\right)$.
The prior probability $p\left(O=(0,1)|s_{1},s_{2}\right)$ is $1/2$,
so the decision rule only depends on the likelihood ratio:

\begin{equation}
\frac{p\left(s_{1},s_{2}|O=(1,0)\right)}{p\left(s_{1},s_{2}|O=(0,1)\right)}=\frac{\lambda\left(s_{1}\right)\varphi\left(s_{2}\right)}{\lambda\left(s_{2}\right)\varphi\left(s_{1}\right)}=\frac{\lambda\left(s_{1}\right)}{\lambda\left(s_{2}\right)}\label{eq:2AFC-likelihood-ratio}
\end{equation}

The optimal decision rule consist in comparing $\lambda(s_{2})$ to
$\lambda\left(s_{1}\right)$, which is equivalent to using $\lambda$
as a saliency map (or any other monotonic transformation of $\lambda$).
The probability of a correct decision (\ref{eq:2AFC-prob-correct})
is then:

\begin{equation}
p_{c}=\int_{\Omega}\lambda(s_{1})\left\{ V^{-1}\int_{\Omega}\I\left(\lambda\left(s_{1}\right)\geq\lambda(s_{2})\right)\mbox{d}s_{2}\right\} \mbox{d}s_{1}\label{eq:2AFC-optimal-uniform}
\end{equation}

The inner integral $\int_{\Omega}\I\left(\lambda\left(s_{1}\right)\geq\lambda(s_{2})\right)\mbox{d}s_{2}$
corresponds to the total volume of the set of all locations with lower
density than the value at $s_{1}$: a contour set of $\lambda$, which
as we have seen is also a minimum volume set. This observation leads
to another of expressing the integral in (\ref{eq:2AFC-optimal-uniform}):
intuitively, each location $s_{1}$ will be on the boundary of a minimum
volume set $F_{\alpha}$, for some value of $\alpha$, and the inner
integral will correspond to the volume of that set. We re-express
the integral by grouping together all values of $s_{1}$ that lead
to the same set $F_{\alpha}$, and hence the same value of the inner
integral: these are the set of values $s_{1}$ that fall along a density
contour $\lambda\left(s_{1}\right)=\lambda_{\alpha}$. Suppose that
we generate a random value of $s_{1}$ and note the $\alpha$ value
of the contour set $s_{1}$ falls on: call $a$ this random variable.
By definition the event $s_{1}\in F_{\alpha}$ happens with probability
$\alpha$, so that $p(a\leq\alpha)=\alpha$, and therefore $a$ has
a uniform distribution over $[0,1]$ (it is in fact a p-value). We
can express Equation (\ref{eq:2AFC-optimal-uniform}) as an expectation
over $a$:

\begin{equation}
p_{c}=\int_{0}^{1}p(a=\alpha)\left(1-V\left(\alpha\right)\right)\mbox{d}a=1-\int_{0}^{1}V\left(\alpha\right)\mbox{d}\alpha\label{eq:AUC-precision}
\end{equation}

Here $V\left(\alpha\right)$ is the relative volume of the minimum-volume
set with confidence level $\alpha.$ $V(0.9)=0.2$ means that, for
a location $s$ sampled from $\lambda$, the smallest region of space
we can find that includes 90\% of the observations takes up just 20\%
of the observation window. For a uniform distribution $V\left(\alpha\right)=\alpha$.
Whatever the density $\lambda$, $V\left(0\right)=0$, and if small
regions include most of the probability mass we are going to see a
slow increase in $V\left(\alpha\right)$ as $\alpha$ rises. This
will lead in turn to a low value for the integral $\int_{0}^{1}V\left(\alpha\right)\mbox{d}\alpha$.
Having small regions that contain most of the probability mass is
the same as having a concentrated or precise point process, and therefore
Equation (\ref{eq:AUC-precision}) shows that under the optimal decision
rule the AUC value measures the precision of the point process.

We now turn to the case where non-fixated locations are not taken
from the uniform distribution: for example, when those locations are
randomly drawn from fixations observed on other pictures. The optimal
rule is again Bayes' rule, Equation \ref{eq:2AFC-likelihood-ratio},
which is equivalent to using $m(s)=\lambda(s)/\varphi\left(s\right)$
as a saliency map. Under an assumption of center bias this may artificially
inflate the AUC value of saliency maps which have low values around
the center. This problem can be remedied by computing an estimate
of the intensity $\varphi$ and using $m(s)/\varphi(s)$ (or more
conveniently $\log m(s)-\log\varphi\left(s\right))$ instead of $m(s)$
when computing AUC scores.

\begin{figure}
\begin{centering}
\includegraphics{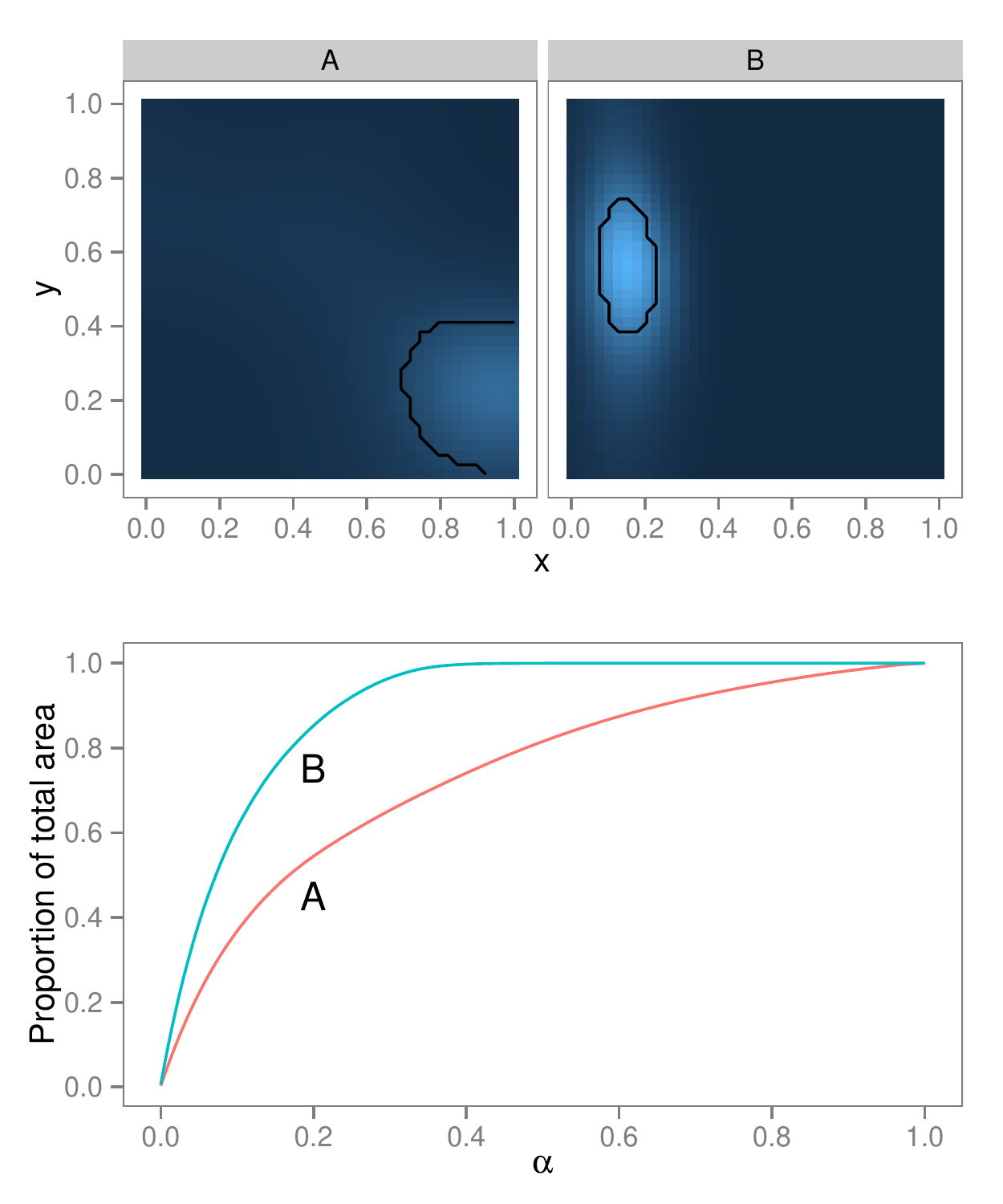} 
\par\end{centering}

\caption{Minimum volume sets. We show in the upper panel two density functions
(A) and (B) along with minimal volume sets with $\alpha=0.8$: these
are the smallest areas containing 80\% of the probability, and correspond
to contours of the density (see text). The $\alpha=0.8$ area is much
larger in A than in B, which reflects higher uncertainty. In the lower
panel, the area of the minimal volume set of level $\alpha$ is represented
as a function of the confidence level $\alpha$. The integral of this
function is shown in the text to equal optimal ROC performance in
the patch classification problem, and reflects the underlying uncertainty
in the point process. \label{fig:Minimum-volume-sets}}
\end{figure}

\paragraph{Area counts}

In area counts the goal is to have a small region that contains as
many fixations as possible. Given a discrete saliency map, we can
build a region that contains the 20\% most salient pixels, and count
the number of fixations that occured there. In this section we show
that if fixations come from a point process with intensity $\lambda$,
the optimal saliency map is again $\lambda$ (again, up to arbitrary
monotonic transformations). In that context, we show further that
if we remove the arbitrariness of setting a criterion at 20\%, and
integrate over criterion position, we recover exactly the AUC performance
measure - the two become equivalent.

Let us define the following optimisation problem: given a probability
density $\pi$, we seek a measurable set $G$ such that

\begin{align}
G_{q} & =\underset{F\in\mathcal{M}\left(\Omega\right)}{\mbox{argmax}}\int_{G}\pi\label{eq:maximum-probability-set}\\
\mbox{s.t}\, & V\left(G\right)=q\nonumber 
\end{align}

that is, among all measurable subsets of $\Omega$ of relative volume
$q$, we seek the one that has maximum probability under $\pi$. These
maximum-probability sets and the minimum-volume-sets defined above
are related: indeed the optimisation problems that define them (\ref{eq:minimum-volume-set}
and \ref{eq:maximum-probability-set}) are dual. This follows from
writing down the Lagrangian of \ref{eq:maximum-probability-set}:

\[
\mathcal{L}\left(G,\eta\right)=\int_{G}\pi+\eta\left(V\left(G\right)-q\right)
\]

which is equivalent to that of (\ref{eq:minimum-volume-set}) for
some value of $\eta$. This result implies that the family of solutions
of the two problems are the same: a maximum-probability set for some
volume $q$ is a minimum-volume set for some confidence level $\alpha$.
Since we know that the family of contours sets is the family of solutions
of the minimum-volume problem, it follows that it is also the family
of solutions of the maximum-probability problem.

In turn, this implies that if fixations come from an IPP with intensity
$\lambda$, the optimal saliency map according to the area count metric
must have the same top 20\% values in the same locations as $\lambda$.
To remove the arbitrariness associated with the criterion, we measure
the total probability in $G_{q}$ for each value of $q$ between 0
and 1 and integrate:

\[
A_{c}=\int_{0}^{1}\left(\int_{G_{q}}\lambda\left(s\right)\mbox{d}s\right)\mbox{d}q=\int_{0}^{1}\mbox{Prob}\left(q\right)\mbox{d}q
\]

The integrand $\mbox{Prob}\left(q\right)$ measures the probability
contained within the maximum-probability set of size $q$: because
of the equivalence of maximum-probability and minimum-coverage sets,
$\mbox{Prob}(q)$ is the inverse of the function $V\left(\alpha\right)$,
which measured the relative size of the minimum-volume set with confidence
level $\alpha$. Therefore $A_{c}=1-\int_{0}^{1}V\left(\alpha\right)\mbox{d}\alpha$,
which is exactly the optimal AUC performance under uniform samples,
as shown in the previous section. Area counts and ROC performance
are therefore tightly related.

\part*{Acknowledgments}

The authors would like to thank Torsten Betz for insightful comments
on the manuscript. This work was funded, in part, by the German Federal
Ministry of Education and Research (BMBF) through the Bernstein Computational
Neuroscience Programs FKZ 01GQ1001F (Ralf Engbert, Potsdam), FKZ 01GQ1001B
(Felix Wichmann, Berlin) and FKZ 01GQ1002 (Felix Wichmann, Tübingen).

 \bibliographystyle{apalike}
\bibliography{ref}

\end{document}